\documentstyle[12pt]{article}

\newcommand{\be}{\begin{equation}}
\newcommand{\ee}{\end{equation}}
\newcommand{\ba}{\begin{eqnarray}}
\newcommand{\ea}{\end{eqnarray}}
\newcommand{\la}{\lambda}
\newcommand{\al}{\alpha}

\newcommand{\tr}{\rm tr}

\newcommand{\e}{\epsilon}

\begin{document}
%\hoffset=-.4truein\voffset=-0.5truein
%\setlength{\baselineskip}{14pt}
\setlength{\textheight}{8.5 in}
\begin{titlepage}
\begin{center}
\vskip 0.6 in
{\large \bf  { Random Matrix,  Singularities  and Open/Close Intersection Numbers
\\}}
\vskip .6 in
\begin{center}
{\bf E. Br\'ezin$^{a)}$}{\it and} {\bf S. Hikami$^{b)}$}
\end{center}
\vskip 5mm
\begin{center}
{$^{a)}$ Laboratoire de Physique
Th\'eorique, Ecole Normale Sup\'erieure}\\ {24 rue Lhomond 75231, Paris
Cedex
05, France. e-mail: brezin@lpt.ens.fr{\footnote{\it
Unit\'e Mixte de Recherche 8549 du Centre National de la
Recherche Scientifique et de l'\'Ecole Normale Sup\'erieure.
} }}\\
{$^{b)}$Okinawa Institute of Science and Technology Graduate University,} 
{1919-1 Tancha, Onna-son, Okinawa, 904-0495, Japan. e-mail:hikami@oist.jp}\\
\end{center}     
\vskip 3mm         

{\bf Abstract}                  
\end{center}
The $s$-point correlation function of a Gaussian  Hermitian random matrix theory, with an external source   tuned to generate a multi-critical singularity,  provides  the intersection numbers of the moduli space for the $p$-th spin curves  through a duality identity.  For one marked point,   the intersection numbers are  expressed to all order in the genus by  Bessel functions.  The matrix models for the Lie algebras of $O(N)$ and $Sp(N)$  provide the intersection numbers of non-orientable surfaces. 
The  Kontsevich-Penner model, and higher $p$-th Airy matrix model with a logarithmic potential, are   investigated for the open intersection numbers, which describe the topological invariants of non-orientable surfaces with boundaries.   String equations for open/closed Riemann surface are derived from the  structure of  the $s$-point correlation functions. The Gromov-Witten invariants of $CP^1$ model are evaluated for one marked point as an application of the present method.
\end{titlepage}
\vskip 3mm

%*******************************

\vskip 3mm
\section{Introduction}
\setcounter{equation}{0}
\renewcommand{\theequation}{1.\arabic{equation}}
\vskip 5mm

The intersection numbers of moduli space of Riemann surface are topological invariants, which are closely related to universal singularities in random matrix models with an external source. 
By  tuning appropriately  the matrix source, multi-critical behaviors are obtained at the edge of the density of state \cite{BH01,BH03,BH04}. They are described by Airy and higher Airy kernels for the $p$-th degenerate singularity. The correlation functions for  the $p$-th singularity  turn out to be generating functions for the intersection numbers of $p$-spin curves on Riemann surfaces \cite{BH3,BH2,BHC3,BH5}.
In this article, we extend our previous work on  the calculation of the intersection numbers of $p$ spin curves. This technique relies on a duality \cite{BH3,BHC3} which is specified in the next section.  

 Recently, the open intersection theory with boundaries has been investigated in \cite{Pandharipande,Buryak1,Buryak2}.  The generating matrix model for these open intersection numbers  is  the  Kontsevich-Penner model \cite{Alexandrov1,Alexandrov2}, namely Kontsevich's model with a logarithmic potential,  studied before in \cite{BH07,MMS}. The Virasoro equations for this case  provide a different structure from KdV, since the Riemann surface with boundaries is no longer orientable.  The expansion of the logarithmic matrix model, like a ribbon graph expansion of Kontsevich matrix model, provides the non-orientable surface. The situation is similar to that of  non-orientable surfaces  generated 
by the matrix models of $O(N)$ and $Sp(N)$ Lie algebras \cite{BH4}. We compute here  the open intersection numbers, and higher $p$-th spin curves, including such Lie algebras.

 The composition of this article  is the following.
In section 2, several dualities  exchanging the size of the random matrices in an external matrix with the number of points in correlation functions, are recalled. The definition of the intersection numbers is briefly recalled. In section 3, the intersection numbers  for $p=2,3,4,5$, and  $p=-1$ (Euler characteristics)  and one marked point, 
are explicitly given  in terms of Bessel functions.
In section  4,  the intersection numbers for  the Lie algebras of  O(2N), O(2N+1), and Sp(N) are discussed. 
The Euler characteristics for those non-orientable surfaces follow.
In
section 5, the  open
intersection numbers  are computed  for one marked point from the Kontsevich-Penner model , and the relation to the $O(N)$ model for non-orientable intersection numbers is discussed.
In section 6,  the intersection numbers for multiple marked points are evaluated and the relation to Virasoro equations is examined. Section 7 is application of the present method to the  Gromov-Witten invariants of $CP^1$.
Section 8 is devoted to  discussions. In an appendix we study the case  of $p$-spin curves for 
open Riemann surfaces.

 %%%%%%%%%%%%%%%%%%%%%%%%%%%%
 \setcounter{equation}{0}
\renewcommand{\theequation}{2.\arabic{equation}}
\vskip 5mm
\section{Dualities}
\begin{itemize}
\item { \bf {GUE ensemble}}

   We have discussed in earlier publications the possibility of computing  topological invariants  relative to Riemann surfaces using a Gaussian ensemble of $N\times N$ Hermitian random matrices with appropriately tuned  external matrix sources $A$. The method relies on two  basic ingredients : i) a totally explicit formula for the K-point correlation functions for arbitrary given source matrix $A$, based on  the HarishChandra-Itzykson-Zuber integral over the unitary group\cite{BH01},  ii) a duality for the correlation functions of K characteristic polynomials $<\prod\limits_1^K \det(\la_{\al} -M)>_A$, with a probability distribution
   \be \label{weight}  P_A(M) = \frac{1}{Z_N} \e^{-\frac{1}{2} \tr M^2 - \tr MA} \ee
 This duality exchanges the size N of the matrices with K, the number of points, i.e.  the $N\times N$ Hermitian random matrices are replaced by $K\times K$ Gaussian random matrices; the   $N\times N$ source matrix A is exchanged with the $K\times K$ source matrix $\Lambda$ whose eigenvalues  are the $\la_{\al}$ : it reads \cite{BH3}
 
\ba \label{dual} 
&&\frac{1}{Z_N} \int d^{N^2} M \prod_{\al =1}^K \det(\la_{\al} -M) \e^{-\frac{1}{2} \tr(M+A)^2}  \nonumber \\&&=(-i)^{NK}     \frac{1}{Z_K} \int d^{K^2} B \prod_{i =1}^N \det(a_i\delta_{\al \beta} -B_{\al \beta})\e^{-\frac{1}{2} \tr(B+i\Lambda)^2} \ea
      
This duality is clearly well adapted to the large $N$ limit since the r.h.s. is an integral over matrices whose size is independent of $N$. But we want  to briefly summarize how else we have used it.

 Tuning appropriately the eigenvalues of the source matrix A, one can obtain in the dual version the Airy matrix model which was introduced by Kontsevich \cite{Kontsevich} ; it appears here as an edge singularity, reminiscent of the Tracy-Widom kernel\cite {TW} which governs the vicinity of the edge of Wigner's semi-circle. This is done by  taking for the source matrix $A$ the identity matrix and  considering the large $N$  scaling regime in which the  $\la_{k}$  are  close to one, namely  $ N^{2/3} (\la_k -1)$ finite, the r.h.s. of (\ref{dual}) becomes the Airy matrix integral introduced by Kontsevich. In this regime one finds 
\ba  \label{Kont}&& \frac{1}{Z_N} \int d^{N^2} M \prod_{\al =1}^K \det(\la_{\al} -M) \e^{-\frac{1}{2} \tr(M+A)^2} \nonumber \\ &&= e^{\frac{N}{2} \tr \Lambda^2} \int d^{K^{2}} B  e^{i \frac{N}{3} \tr B^3 +iN \tr B(\Lambda -1)} \ea
and the r.h.s. after a rescaling $B \to B N^{-1/3}$ , $(\la-1) \to (\la-1)N^{-2/3} $ reduces to  Kontsevich Airy integral ; 
\be Z_{KT}= \int dB e^{ \frac{i}{3} \tr B^3 - \tr \Lambda B^2} \ee 
In the r.h.s. of (\ref{Kont}) tuning the $a_{\alpha}$  one may generate the Airy matrix integral, whereas the l.h.s. is still a Gaussian integral whose correlation functions are known explicitly. 
Indeed the one-point function with the probability weight (\ref{weight})  is given by \cite{BH3}
\be U(\sigma) = \langle \tr e^{\sigma M}\rangle =  \frac{1}{\sigma} \oint \frac {du}{2i\pi} e^{\sigma u} \prod_1^N \frac{u- a_{\al} +\sigma}{u- a_{\al} } \ee
in which the $a_{\al}$ are the eigenvalues of the source matrix $A$. This formula is exact for any N.  For the s-point function \cite{BH3} the result is an integral over s complex variables
\ba \label{10} &&U(\sigma_1\cdots \sigma_s) = \langle \tr e^{\sigma_1M}\cdots \tr e^{\sigma_s M}\rangle \nonumber\\
&&= e^{\frac{1}{2} \sum_1^s \sigma_i^2} \oint \prod_1^s \frac {du_i e^{\sigma _i u_i}}{2i\pi} \det \frac{1}{u_i+\sigma_i-u_j} \prod_{i=1}^s\prod_{\al =1}^N  (1 + \frac {\sigma_i}{ u_i-a_{\al}}) \nonumber\\
\ea 
 Let us illustrate on the simplest example how one can use (\ref {Kont}) and (\ref{10}),  for the one-point function. We take $\la_a = \la$ for $a= 1 \cdots K$. Then
we are dealing with
\be < [\det (\la -M)]^K> _A = < [ \det (1-iB)]^N >_{\Lambda} \ee
and make  use of "replicas", i.e.  of the identity 
\be \lim_{K\to 0}\frac{1}{K} \frac {d}{d\la }[\det (\la -M)]^K = \tr \frac {1}{\la -M} \ee
Since $\la$ is in the vicinity of the edge of Wigner's  semi-circle, the resolvent has to be computed in this regime, but knowing explicitly $U(\sigma)$, this is straightforward. 

The intersection numbers of moduli space of curves are defined as coefficients in the expansion
of $t_n = {\rm tr}\frac{1}{\Lambda^{n+\frac{1}{2}}}$ for $Z_{KT}$. The coefficients
, the intersection numbers $<\tau_{n_1}\cdots \tau_{n_s}>_{g}$ , are defined also as
\be
<\tau_{n_1}\cdots \tau_{n_s}>_g = \int_{\overline M_{g,s}} \psi_1^{n_1}\cdots \psi_s^{n_s}
\ee
 where $\psi_i$ is called as $\psi$ class and equal to $c_1({\mathcal{L}}_i)$ with $c_1$ first Chern class and ${\mathcal{L}}_i$ is line bundle at marked point $i$.  We have shown before
 that $U(\sigma_1,...\sigma_s) =<{\rm tr} e^{\sigma_1 M}\cdots \tr e^{\sigma_s M}>$ is generating function of the intersection numbers, since it is a Fourier transform of the density correlation functions,
 \be
 U(\sigma_1,...\sigma_s) = <\int \prod_i^s d\lambda_i e^{\sigma_i \lambda_i} \tr \delta(\lambda_i - M)>_A
 \ee
 This function provides a polynomial expansion of $\sigma_i$. The degree of total $ \sigma_i$
 is equal to $\sum_i^s (n_i + \frac{1}{2})$, and $\sum_i^s n_i = 3 g - 3 + s$. This is similar
 to the evaluation of the intersection numbers by hyperbolic surfaces \cite{Mirzakhani}, by replacing the parked points by disks, whose perimeter lengths are $l_1,....,l_s$ and the generating function of the intersection numbers is a polynomial of $l_i$, (i=1,...,s) and the total degree is $6g - 6 + 2 s$ \cite{Do}.
 
We have shown earlier, using this strategy together with replicas, how to compute from there
the intersection numbers of the moduli of curves on Riemann surfaces
with one marked point \cite{BH3} ; the method clearly allowed for more marked points. 
Higher multi-critical singularities, characterized by an integer $p$ may also  be tuned from appropriately chosen external sources $A$ \cite{BH03,BH04}.  One  may obtain thereby  a generalized p-th Airy matrix model \cite{BHC3}, taking
 $A={\rm diag}(a_1,...,a_1,....,a_{p-1},....,a_{p-1})$, with $(p-1)$ distinct  eigenvalues values, each of them being ($\frac{N}{p-1}$) times degenerate.
The  conditions\cite{BH03} are
 \be\label{condition}
 \sum_{\alpha=1}^{p-1} \frac{1}{a_\alpha^2} = p-1, \hskip 3mm \sum_{\alpha=1}^{p-1}
 \frac{1}{a_\alpha^m} = 0 \hskip 2mm(m=3,...,p), \hskip 2mm\sum_{\alpha=1}^{p-1}
 \frac{1}{a_\alpha^{p+1}} \neq 0
 \ee
and  we obtain the p-th degenerated Airy matrix model. The  correlation functions are known in integral form 
 \ba
&& U(\sigma_1,...,\sigma_n) \nonumber\\
&&=\oint \prod \frac{d u_i}{2\pi}  e^{-\frac{N}{p^2 -1} \sum_{\alpha=1}^{p-1} \frac{1}{a_\alpha^{p+1}}[(\sum_{i=1}^n (u_i + \frac{1}{2N}\sigma_i)^{p+1} - \sum_{i=1}^n (u_i - \frac{1}{2N }\sigma_i)^{p+1})]} \nonumber\\
 &&\times\prod {\rm det}( \frac{1}{u_i - u_j  + \sigma_i})
 \ea
 The one point function, which corresponds to one marked point, becomes in the scaling limit
 \be\label{U1}
 U(\sigma) = \frac{1}{N\sigma} \int \frac{du}{2\pi i} e^{-\frac{c}{p+1}[(u+\frac{1}{2}\sigma)^{p+1} - (u - \frac{1}{2}\sigma)^{p+1}]}
 \ee
 with $c = \frac{N}{p-1}\sum_{\alpha=1}^{p-1}\frac{1}{a_\alpha^{p+1}}$.
The  Airy matrix model corresponds to the $p=2$ case. It is also possible to continue the model to negative values of $p$. In particular  the case $p=-1$ provides a generating function for the orbifold Euler characteristics of surfaces with $n$ marked points, allowing us to recover through this method the classic results of \cite{Harer,Penner, BHC3}.

\item {\bf{ Lie algebras of classical groups} }

The previous  duality (\ref{dual}) extends  to Lie algebras of the classical groups, such as antisymmetric real matrices for the orthogonal group $O(2N)$. \be\label{eq1}
< \prod_{\alpha=1}^k {\rm det}(\lambda_\alpha\cdot {\rm I} - X )>_A
= <\prod_{n=1}^N {\rm det}( a_n\cdot {\rm I} - Y ) >_\Lambda
\ee
where $X$ is $2N\times 2N$ real antisymmetric matrix ($X^{t} = - X$ ) 
and $Y$ is $2k\times 2k$ real 
antisymmetric matrix ;  the eigenvalues of $X$ and $Y$ are thus pure imaginary. $A$ is also a $2N \times 2N$ antisymmetric matrix, and it couples to
X as an external matrix source.  The matrix $\Lambda$ is $2k\times 2k $ antisymmetric matrix, coupled to $Y$.
We assume, without loss of  generality, that $A$ and $\Lambda$ have the  canonical form : 

\be\label{canonical}
A = \left( \matrix{ 
0 & a_1 & 0 & 0&\cdots\cr
-a_1
&0& 0& 0&\cdots\cr
0&0&0& a_2&0\cr
0&0&-a_2&0&0\cr
\cdots} \right),
\ee
i.e.
\be
A = a_1 v \oplus  \cdots \oplus a_N v, \hskip 5mm v =i\sigma_2 = \left( \matrix{0&1\cr -1&0}\right ).
\ee
$\Lambda$ is expressed also as
\be
\Lambda = \lambda_1 v \oplus \cdots \oplus \lambda_k v.
\ee
The characteristic polynomial $ {\rm det}(\lambda\cdot {\rm I}-X)$ has the $2N$ roots,
 
$(\pm i\la_1, \cdots, \pm i\la_n)$. 
The Gaussian averages in  (\ref{eq1}) are defined as 
\ba
< \cdots>_A &=& \frac {1}{Z_A}\int dX e^{\frac{1}{2}{\tr} X^2 + {\tr X A}}\nonumber\\
<\cdots>_\Lambda &=&\frac{1}{Z_{\Lambda}} \int dY e^{\frac{1}{2}{\tr}Y^2 + {\tr Y \Lambda}}
\ea
in which $X$ is a $2N\times2N$ real antisymmetric matrix, and $Y$ a $2k\times2k$ real antisymmetric matrix ; the coefficients $Z_A$ and $Z_{\Lambda}$ are such that the expectation values of one is equal to one. The derivation relies on a representation of the characteristic polynomials in terms of integrals over Grassmann variables,  as  for the  $U(N)$  case, but it is more involved \cite{BH4}. 

Here again the Harish Chandra formula leads to explicit formulae for the correlation functions. The one-point function for instance is 
\be
U(s) = -\frac{1}{N s} \oint \frac{dv}{2\pi i}\prod_{n=1}^N \left( \frac{v^2+ a_n^2}{
(v+ \frac{s}{2})^2 + a_n^2}\right)
\frac{v + \frac{s}{2}}{v+ \frac{s}{4}}e^{v s + \frac{s^2}{4}},
\ee
and higher point functions are also known explicitly. Therefore one may repeat the same tuning plus duality strategy in this case, leading to the desired topological numbers for non-orientable surfaces generated by these antisymmetric matrix models. 

\item {\bf{ Superduality}}
\vskip 2mm
 
 Consider 
 \be
 \label{superduality}
   F_{P,Q} (\lambda_{\alpha}\cdots \mu_{\beta}\cdots) = \frac{1}{Z_N}< \frac{\prod_{\alpha=1}^P {\rm det}(\lambda_\alpha - M) }{\prod_{\alpha=1}^Q {\rm det}(\mu_\beta - M) }>_{A}\ee
 with
 \be \langle {\cal{O}}(M)\rangle_A = \frac{1}{Z_A} \int dM {\cal{O}}(M)e^{-\frac{1}{2} \tr M^2 + \tr MA} \ee
 For instance the average resolvent is given by $P=Q=1$ , after taking derivative with respect to $\lambda$ and  setting $\lambda = \mu$. 
 
 Let us recall standard definitions for supermatrices: let
\be X = \ ( \begin{array}{clcr} a &\al \\ \beta &b \end{array} \ ) \ee
in which the matrix elements of  $a$ and $b$ are commuting numbers, those of $\alpha$ and $\beta$ anticommuting.  Then the supertrace
\be {\rm{str}} X = {\rm{tr}} a- {\rm{tr}} b \ee
ensures the cyclic invariance. The superdeterminant  is given by
\be \label{sd} {\rm{sdet}}X = \frac{ \det a}{ \det (b-\beta a^{-1} \al)} = \frac{\det ( a-\al b^{-1} \beta)}{\det b} \ee
 based on the integral
\be \int  d\theta d\bar\theta   dx d\bar{x} \e^{i {\overline{ \Phi}} X \Phi} = ({\rm{sdet}} X)^{-1} \ee
where 
\be \Phi = \ ( \begin{array}{clcr} x  \\ \theta \end{array} \ ) \hskip 1cm  \overline\Phi = \ ( \begin{array}{clcr} \bar x  &\bar \theta \end{array} \ )\ee
The formulae are obtained either by integrating first the commuting variables, or  the anticommuting variables first.
We use the conventions
\be \overline{\theta_1\theta_2} =   \bar{\theta_1}\bar {\theta_2} \ee
and
\be \overline{\overline \theta} = \theta \ee

Finally the usual bosonic formula still holds here, namely
\be {\rm {str}}(\log X )= \log {({\rm {sdet}X})}. \ee

We are now in position to derive the duality formula for (\ref{F}) which we first write in integral form as 

 \ba \label{F} F_{P,Q} (\lambda_{\alpha}\cdots \mu_{\beta}\cdots) = \int \prod_{a=1}^N  \prod_{\alpha =1}^ Q\prod_{\beta =1}^P d\bar x^a_{\alpha} dx^a_{\alpha}d\bar{\theta}^a_{\beta} d{\theta}^a_{\beta} \nonumber \\\langle e^{-\sum_{\alpha =1}^P \bar x_{\alpha}^a (\lambda_{\alpha} -M)_{ab} x_{\alpha}^b -\sum_{\beta =1}^Q \bar \theta_{\beta}^a(\mu_{\beta} -M)_{ab} \theta_{\beta}^b}\rangle_A \ea
 or, introducing the $(Q+P)\times (Q+P)$ diagonal matrix $\Lambda$ made of $\mu_{\beta}$ , $\beta = 1 \cdots Q$ and $\lambda_{\alpha}$ , $\alpha =1\cdots P$ 
  \be  F_{P,Q} (\lambda_{\alpha}\cdots \mu_{\beta}\cdots) = \int \prod_{a=1}^N d\bar x^a_{\alpha} dx^a_{\alpha}d\bar{\theta}^a_{\beta} d{\theta}^a_{\beta}\langle e^{-\ \bar\Phi^a \Lambda \Phi^a +\bar\Phi^a \cdot M_{ab}\Phi^b }\rangle_A \ee
  Since
  \be \langle e^{\tr XM} \rangle_A = e^{\frac{1}{2} \tr X^2+ \tr AX} \ee
we have
 \be X_{ba} = \bar\Phi^a \cdot \Phi^b = \sum_{\alpha=1}^{P+Q}\bar \Phi^a _{\alpha} \Phi^b_{\alpha} .\ee
 
 Then 
 \be \tr (AX) = \sum_{n=1}^N a_n \sum_{\alpha=1}^{P+Q}\bar \Phi^n _{\alpha} \Phi^n_{\alpha}  \ee
 in which the $a_n$ are the eigenvalues of $A$,
 \be \tr X^2 =\sum_{a,b =1}^N  \sum_{\alpha,\beta=1}^{P+Q}\bar \Phi^a _{\alpha} \Phi^b_{\alpha} \bar \Phi^b _{\beta} \Phi^a_{\beta} .\ee
 Let us define the matrix $\Gamma$, $(Q+P)\times (Q+P)$
 \be \Gamma_{\alpha,\beta} = \sum_{a =1}^N \overline{ \Phi^a _{\alpha} }\Phi^a_{\beta} =  \biggl( \begin{array}{clcr} \Gamma_1&\Gamma_2 \\ \Gamma_2^{\dagger}& {\Gamma_3} \end{array} \biggr)= \biggl(\begin{array}{clcr}  \bar x\cdot x&\bar\theta \cdot x \\\bar x\cdot \theta & \bar \theta \cdot \theta  \end{array} \biggr)\ee
 
 This matrix $\Gamma_1$ is Hermitian but $\Gamma_3$ is anti-Hermitian. 

  To express $\tr X^2$ in terms of the matrix $\Gamma$ some commutations are required and 
  one obtains easily
  \be \tr X^2 = \sum_{\alpha,\beta}=s\tr (\Gamma^2) \ee
  Therefore
  \ba  F_{P,Q} (\lambda_{\alpha}\cdots \mu_{\beta}\cdots) &&= \int \prod_{a=1}^N d\bar x^a_{\alpha} dx^a_{\alpha} d\bar{\theta}^a_{\beta} d{\theta}^a_{\beta} \nonumber\\
  &&\ e^{-\ \bar\Phi^a \Lambda \Phi^a + \sum_{n=1}^N a_n \sum_{\alpha=1}^{P+Q}\bar \Phi^n _{\alpha} \Phi^n_{\alpha}  +\frac{1}{2} s\tr \Gamma^2} \ea
  The SUSY Hubbard-Stratonovich transformation reads 
  \be \int d\Delta e^{s\tr (-\frac{1}{2}  \Delta^2 + \Delta \Gamma)} = e^{\frac{1}{2} s\tr \Gamma^2} \ee
  in which $\Delta$ is  $(P+Q)\times(P+Q)$ and like $\Gamma$ as far as hermiticity is concerned.
  Then 
   \ba && F_{P,Q} (\lambda_{\alpha}\cdots \mu_{\beta}\cdots) =\int d\Delta  \int \prod_{a=1}^Nd\bar x^a_{\alpha} dx^a_{\alpha}d\bar{\theta}^a_{\beta} d{\theta}^a_{\beta} \nonumber\\
   &&\ e^{-\ \bar\Phi^a \Lambda \Phi^a + \sum_{n=1}^N a_n \sum_{\alpha=1}^{P+Q}\bar \Phi^n _{\alpha} \Phi^n_{\alpha} }  
  e^{ -\frac{1}{2} s\tr \Delta^2 + s\tr \Delta \Gamma}   \ea
 
 One can integrate out on the $x$'s and $\theta$'s. The quadratic form in the exponential is
 $$ - \bar\Phi^a \Lambda \Phi^a +  \sum_{n=1}^N a_n \sum_{\alpha=1}^{P+Q}\bar \Phi^n _{\alpha} \Phi^n_{\alpha} + \Delta_{\alpha, \beta}\bar \Phi^a _{\beta}  \Phi^a_{\al} (-1)^{F_{\beta}} $$
 in which $F_{\beta} = 0$ for $1\leq \beta\leq P$ or $F_{\beta} = 1$ for $(P+1)\leq \beta\leq (P+Q)$
 The integration then gives
 $$\prod_1^N s\det\ ^{-1}[ ( \Lambda_{\al} - a_n) \delta_{\al \beta} -  \Delta_{\al \beta}(-1)^{F_\beta} ]$$
 Therefore  we change  $ \Delta_{\al \beta}(-1)^{F_\beta}\to \tilde \Delta_{\al \beta}$ and one verifies that 
 \be s\tr \Delta^2 = s\tr \tilde\Delta^2\ee.  Then one ends up with
 \be {\label{sdual}} F_{P,Q} (\lambda_{\alpha}\cdots \mu_{\beta}\cdots)  = \int d\Delta e^{-\frac{1}{2} s\tr \Delta^2} \prod_1^N s\det\ ^{-1}[ ( \Lambda_{\al} - a_n) \delta_{\al \beta} -  \Delta_{\al \beta}] \ee

The above identity (\ref{sdual}) relates an ordinary integral to a super matrix integration. In this sense it is not a full duality although it can be used for the large N-limit or for a super-generalization of the Kontsevich model.  However a full  superduality has been derived  by Desrosiers and Eynard for expectation values of ratios of super-determinants \cite{D-E} and our identity appears as a simple limiting case.  

\vskip 10mm

\item {\bf{ Arbitrary $\beta$}}\\
 An extension of the GUE duality (\ref{dual}) to the three classical Gaussian ensembles GOE, GUE, GSE with respectively  $\beta= 1,2,4$ has been derived by Desrosiers \cite{D} , but it exchanges $\beta$ to $4/ \beta$. However the lack of HarishChandra formula for integrating over the orthogonal or symplectic group does not allow one to compute explicitly the $k$-point functions and we cannot repeat the steps that we have followed for $\beta = 2$. However we have used supergroup methods to obtain the one and two-point functions \cite{ BH2,BH6,BHL}. 
\end{itemize}

%%%%%%%%%%%%%%%%%%%%%%%%%%%%%%%
 %%%%%%%%%%%%%%%%%%%%%%%%%%% 
\section{ GUE }
\setcounter{equation}{0}
\renewcommand{\theequation}{3.\arabic{equation}}
\vskip 5mm
\underline{ $X \in U(N)$}
\vskip 2mm

We now use the duality formula  for  computing the one-point function $U(\sigma)$ for some
symmetric spaces.\\
(i)  Let us first consider  the Hermitian case:

From the duality formula   (\ref{superduality}), we obtain 
\ba
F_N(\lambda,\mu) &=& <\frac{{\rm det}(\lambda - X)}{{\rm det}(\mu - X)}>_A\nonumber\\
&=&
\delta_{\lambda,\mu} + N (\lambda - \mu) \frac{e^{-N(\mu^2 - \lambda^2)}}{2\pi^2} \int_{-\infty}^\infty \int_{-\infty}^\infty dt du \prod_{\alpha=1}^N (\frac{a_\alpha - i t}{a_\alpha + u}) \frac{1}{u - it} \nonumber\\
&\times& e^{- \frac{N}{2}u^2 - \frac{N}{2}t^2 - i N t\lambda + N u \mu}
\ea

 The density of state $\rho(\lambda)$ 
 is \cite{BH6}
 \ba
 \rho(\lambda) &=& - \lim_{\mu\to \lambda} \frac{1}{\pi N}\frac{\partial}{\partial \mu} 
 {\rm Im} F_N(\lambda,\mu)\nonumber\\
 &=& \frac{1}{N}\int \frac{dt}{2\pi i}\oint \frac{du}{2\pi i} \prod_{\alpha=1}^N(\frac{a_\alpha - i t}{a_\alpha + u})\frac{1}{u - it}e^{-\frac{N}{2}u^2 -\frac{N}{2}t^2 - i N t \lambda + N u \lambda}
 \ea
 By tuning the external source $a_\alpha$ as (\ref{condition}), and taking Fourier transform of $\rho(\lambda)$,
 \be\label{one}
 U(\sigma) = \frac{1}{N\sigma} \oint \frac{du}{2\pi i} e^{-\frac{c}{p+1}[(u+\frac{1}{2}\sigma)^{p+1} - (u - \frac{1}{2}\sigma)^{p+1}]}
 \ee
which is identical to the previous expression (\ref{U1}).  
  
\vskip 2mm
\noindent
(1)  p=2
\vskip 2mm
The integral (\ref{one}) becomes Gaussian (c=1),
\ba
U(\sigma) &=& \frac{1}{\sigma} e^{-\frac{\sigma^3}{12}}\int_{-\infty}^\infty \frac{du}{2\pi} e^{-\sigma u^2}\nonumber\\
&=& \frac{1}{2\pi \sigma}\sqrt{\frac{\pi}{\sigma}}e^{-\frac{1}{12}\sigma^3}
\ea
This may be expressed as a  modified Bessel function $K_{\frac{1}{2}}(z)$  
\be
K_{\frac{1}{2}}(z) = \sqrt{\frac{\pi}{2z}}e^{- z}
\ee
and we have
\be
U(\sigma) = \frac{1}{2\pi \sqrt{6}} K_{\frac{1}{2}}(\frac{\sigma^3}{12})
\ee
This explicit representation gives the intersection numbers fro Riemann surfaces of genus $g$,
\be
<\tau_{m}>_g = \frac{1}{(24)^g g!}, \hskip 3mm ( m= 3g -2 )
\ee

\vskip 2mm
\noindent
(2)  p=3
\vskip 2mm
Then 
\be
U(\sigma) = \frac{1}{\sigma}\oint \frac{du}{2\pi i}e^{-\sigma u^3 - \frac{1}{4}\sigma^3 u}
\ee
or changing $u$ to  $u = v^{1/3}$,
\be
U(\sigma) = \frac{1}{3\sigma}\oint \frac{dv}{2\pi i} v^{-\frac{2}{3}}e^{- \sigma v - \frac{1}{4}\sigma^3 v^{\frac{1}{3}}}
\ee
The path integral may be   divided into two integrals on the real axis, above and below the cut : 
\ba
U(\sigma) &=& U_I (\sigma) + U_{II} (\sigma)\nonumber\\
U_{I}(\sigma) &=& \frac{1}{3\sigma i}\int_0^\infty \frac{dv}{2\pi}(e^{2\pi i} v)^{-\frac{2}{3}}e^{- \sigma v - \frac{1}{4}\sigma^3 (e^{2\pi i} v)^{\frac{1}{3}}}\nonumber\\
U_{II}(\sigma) &=& - \frac{1}{3\sigma i}\int_0^\infty \frac{dv}{\pi}(e^{-2\pi i} v)^{-\frac{2}{3}}
e^{-\sigma v - \frac{1}{4}\sigma^3 (e^{-2\pi i}v)^{\frac{1}{3}}}
\ea
$U_{II}$ is the complex conjugate of $U_{I}$, and $U(\sigma)$ is  real.
The integer powers of $\sigma$, i.e. $\sigma^n$, cancel.
This corresponds to the spin $j=2$, since
$\sigma^{n + \frac{1+j}{3}} = \sigma^{n+1}$. This cancellation means that there is no Ramond term in $U(\sigma)$, and only Neveu-Schwarz types exist.

We have for $p=3$  
\ba\label{exp}
 U(\sigma) &=& (\frac{\sin \frac{\pi}{3}}{\pi}) [  \frac{1}{3\sigma^{4/3}} \Gamma(\frac{1}{3}) + \frac{1}{12} \sigma^{\frac{4}{3}}\Gamma(\frac{2}{3}) - \frac{1}{3^3\cdot 2^7}\sigma^{\frac{20}{3}}\Gamma(\frac{1}{3}) 
+  ...  \hskip 2mm]\nonumber\\
&=& \frac{1}{6\sqrt{3}}[ J_{\frac{1}{3}}(\frac{1}{12\sqrt{3}}\sigma^4) + J_{-\frac{1}{3}}(\frac{1}{12\sqrt{3}}\sigma^4)]
\ea
This may also be written as an Airy function $Ai(z)$,
\be
U(\sigma) = \frac{1}{3^{\frac{1}{3}}\sigma^{\frac{4}{3}}}Ai( - \frac{1}{4\cdot 3^{1/3}}\sigma^{\frac{8}{3}})
= \frac{1}{\sigma} \int_{-i\infty}^{i \infty} \frac{du}{2\pi i}e^{-\sigma u^3 - \frac{1}{4}\sigma^3 u} = \frac{1}{\sigma}\int_{-\infty}^\infty \frac{du}{2\pi} e^{i \sigma u^3 - \frac{i}{4}\sigma^3 u}
\ee
In deriving (\ref{exp}) we have made use of 
\be
\int_0^\infty \cos (t^3 - x t) dt = \frac{\pi}{3}\sqrt{\frac{x}{3}}[ J_{1/3}(\frac{2 x^{3/2}}{3^{3/2}}) + J_{- 1/3}(\frac{2 x^{3/2}}{3^{3/2}}) ]
\ee
The Airy function $Ai(z)$ was used for the case of two marked points for $p=3$ in \cite{BH5}.

We obtain thus the explicit expression for  the intersection numbers
\be\label{p3}
<\tau_{n,j}>_g = \frac{1}{(12)^g g!} \frac{\Gamma(\frac{g+1}{3})}{\Gamma(\frac{2-j}{3})}
\ee
with $n= (8g-5-j)/3$.  This condition comes from general constraint for the intersection numbers of s-marked points of the moduli spaces of p-spin curves \cite{Witten1},
\be\label{generalcondition}
(p+1)(2g - 2 + s) = p \sum_{i=1}^s n_j + \sum_{i=1}^s j_i + s
\ee
The result of (\ref{p3}) agrees with (\ref{pexpression}) for $p=3$. We have 
\ba
<\tau_{1,0}>_{g=1} &=& \frac{1}{12},\hskip 3mm <\tau_{3,2}>_{g=2} = 0,\nonumber\\
<\tau_{6,1}>_{g=3} &=& \frac{1}{31104},\hskip 3mm <\tau_{9,0}>_{g=4} = \frac{1}{746496}, \hskip 2mm .....
\ea

\vskip 2mm
\noindent
(3)  p=4
\vskip 2mm
\ba
U(\sigma) &=& \frac{1}{\sigma} e^{- \frac{1}{80} \sigma^5} \oint \frac{du}{2\pi i}
e^{- \sigma u^4 - \frac{1}{2}\sigma^3 u^2}\nonumber\\
&=& \frac{1}{4\sigma^{5/4}}e^{- \frac{1}{80}\sigma^5}\oint \frac{dx}{2\pi i}
x^{-\frac{3}{4}} e^{-x - \frac{1}{2}\sigma^{5/2} x^{1/2}}\nonumber\\
&=& \frac{\sin \frac{\pi}{2} }{4\pi \sigma^{5/2}}e^{-\frac{1}{80}\sigma^5}\int_0^\infty dx x^{- 3/4} e^{- x + \frac{1}{2}\sigma^{5/2} x^{1/2}}
\ea
where the contour integral reduces to tow integrals above and below the cut,  as for the $p=3$ case.
We thereby obtain for $p=4$,
\ba
U(\sigma) &=& \frac{1}{4\pi }\sigma^{- 5/4}e^{-\frac{1}{80} \sigma^5}
\sum_{n=0}^\infty \frac{1}{n!}(\frac{\sigma^{5/2}}{2})^n \Gamma(\frac{n}{2}+\frac{1}{4})
\nonumber\\
&=&\frac{1}{4\pi }\sigma^{- 5/4}e^{-\frac{1}{80} \sigma^5} [ 
\Gamma(\frac{1}{4}) + \frac{1}{2}\sigma^{5/2} \Gamma(\frac{3}{4}) + \frac{1}{32} \sigma^5 \Gamma(\frac{1}{4}) +  \cdots]
\ea
This one-point function is the generating function
\be
U(\sigma) = \frac{1}{\pi} \sum_{n,j} <\tau_{n,j}>_{g} \sigma^{n + \frac{1+j}{4}} 4^{g-1} \Gamma(
1 - \frac{1+j}{4})
\ee
with $n = \frac{1}{4}(10 g - 6 - j)$.
Therefore the  intersection numbers  for $p=4$ are 
\ba
&&<\tau_{1,0}>_{g=1} = \frac{1}{8},\hskip 3mm <\tau_{3,2}>_{g=2} = \frac{3}{2560},
\hskip 3mm <\tau_{6,0}>_{g=3} = \frac{3}{20480}, \nonumber\\
&& <\tau_{8,2}>_{g=4} = \frac{77}{39321600}, \hskip 3mm <\tau_{11,0}>_{g=5} = \frac{19}{104857600}, ...
\ea

The exponent of the integrand (\ref{one}) may be expressed as the Chebishev function $T_4(t,x)= t^4 + 4 x t^2 + 2 x^2$ with $x = \frac{1}{8}  \sigma^{5/2}$, and we obtain a closed formula from the result of Appendix IV,
\be\label{analytic}
U(\sigma) = \frac{1}{2\sqrt{8}}e^{\frac{3 }{160}\sigma^5} \frac{1}{2 \sin(\frac{\pi}{4})}
[ I_{-\frac{1}{4}}(\frac{1}{32}\sigma^5) + I_{\frac{1}{4}}(\frac{1}{32}\sigma^5) ]
\ee
Expanding above expression of the modified Bessel function $I_\nu(z)$ for small $\sigma$, we have
\ba\label{p4}
U(\sigma) &=&\frac{1}{8}\sum_{m,n=0}^\infty \frac{1}{m! n! \Gamma(n+\frac{3}{4})}
(\frac{3}{160})^m (\frac{1}{64})^{2n -\frac{1}{4}}\sigma^{5m+10n -\frac{1}{4}}\nonumber\\
&+& \frac{1}{8}\sum_{m,n=0}^\infty \frac{1}{m! n! \Gamma(n + \frac{5}{4})} (\frac{3 }{160})^m (\frac{1}{64})^{2n+\frac{1}{4}}\sigma^{5m + 10 n + \frac{1}{4}}\nonumber\\
&=& \frac{1}{8\pi} \Gamma(\frac{3}{4}) \sigma^{\frac{5}{4}} + \frac{3}{640\pi} \Gamma(\frac{1}{4}) \sigma^{\frac{15}{4}} + \cdots
\ea

with $n = \frac{1}{4}(10 g - 6 - j)$.
which agrees with the previous  results \cite{BHC3} and also it agrees with the result of Liu and Xu derived by the recursion formula from Gelfand-Dikii equation \cite{LiuXu}. We have obtained a closed analytic  formula for the intersection numbers of p=4 spin curves of one marked point in (\ref{analytic}) for arbitrary genus $g$ by expressing it as a Bessel function.
\vskip 2mm
\noindent
(4)  $p=5$
\vskip 2mm
 We have ,
\ba\label{U5}
 U(\sigma) &=& \frac{1}{\sigma}\oint \frac{du}{2\pi i}  e^{-\frac{1}{6}[(u+\frac{\sigma}{2})^6 - (u-\frac{\sigma}{2})^6]}\nonumber\\
&=& \frac{1}{\sigma}\oint \frac{du}{2\pi i} e^{-\sigma u^5 - \frac{5}{6}\sigma^3 u^3 - \frac{1}{16}\sigma^5 u}\nonumber\\
&=& \frac{1}{5 \sigma^{6/5}}\oint \frac{dx}{2\pi i} x^{- 4/5} e^{- x - \frac{5}{6}\sigma^{12/5}x^{3/5} - \frac{1}{16}\sigma^{24/5}x^{1/5}}
\ea
By taking paths around a cut, similar to $p=3,4$ cases, we have
\ba
U(\sigma) &=&\frac{1}{5 i \sigma^{6/5}}e^{-\frac{8\pi i}{5}}\int_0^\infty \frac{dx}{2\pi} x^{-4/5} e^{- x \frac{5}{6}\sigma^{12/5}e^{6\pi i/5}x^{3/5} - \frac{1}{16}\sigma^{24/5}e^{2\pi i/5}x^{1/5}}\nonumber\\
&-& \frac{1}{5 i \sigma^{6/5}}e^{\frac{8\pi i}{5}}\int_0^\infty \frac{dx}{2\pi} x^{-4/5} e^{- x \frac{5}{6}\sigma^{12/5}e^{-6\pi i/5}x^{3/5} - \frac{1}{16}\sigma^{24/5}e^{- 2\pi i/5}x^{1/5}}\nonumber\\
&=& \frac{\sin\frac{2\pi}{5}}{5\pi} \Gamma(\frac{1}{5}) \sigma^{- 6/5} - \frac{\sin\frac{2\pi}{5}}{6\pi} \Gamma(\frac{4}{5}) \sigma^{6/5} - \frac{11 \sin(\frac{\pi}{5})
}{720 \pi} 
\Gamma(\frac{2}{5}) \sigma^{18/5} \nonumber\\
&+& \frac{\sin \frac{\pi}{5}}{\pi} \frac{341}{207360} \Gamma(\frac{3}{5}) \sigma^{\frac{42}{5}} + \cdots\nonumber\\
\ea

We obtain
\ba
&&<\tau_{1,0}>_{g=1}= \frac{1}{6}, \hskip 2mm <\tau_{3,2}>_{g=2} = \frac{11}{3600}\hskip 3mm
<\tau_{5,4}>_{g=3} = 0,\nonumber\\
&&\hskip 3mm <\tau_{8,1}>_{g=4} = \frac{341}{25920000}, \hskip 3mm
<\tau_{10,3}>_{g=5} = \frac{161}{777600000}, ....
\ea
which agrees with \cite{BHC3} and \cite{Liu}.

We  use $u= {\rm sinh} \theta$, and note that $T_5(iu) = i {\rm cosh} 5 \theta$,
\be
U(\sigma) = \sqrt{\frac{2}{3}}\int_0^\infty d\theta \hskip 1mm {\rm cosh}\theta \hskip 1mm {\rm exp}[ - 2 x^{\frac{5}{2}}
{\rm cosh} 5 \theta +\frac{11\sqrt{2}}{16} \sigma^6 {\rm sinh}\theta ]
\ee
with $x = \frac{1}{3}(\frac{1}{32})^{\frac{1}{5}} \sigma^{\frac{12}{5}}$.
By the change of $\theta \to \frac{1}{5}\theta$, we have
\be\label{p5}
U(\sigma) = \frac{1}{5} \sqrt{\frac{2}{3}}\int_0^\infty d\theta e^{- 2 x^{\frac{5}{2}}{\rm cosh} \theta}
\sum_{n=0}^\infty \frac{1}{n!}(\frac{11\sqrt{2}}{16}\sigma^6 {\rm sinh} \frac{\theta}{5})^n {\rm cosh}\frac{\theta}{5}
\ee
This integral is evaluated by the formula,
\be\label{formula2}
\int_0^\infty d\theta e^{-z {\rm cosh}\theta - \nu \theta} = \frac{1}{\sin \nu\pi}\int_0^\pi d \theta e^{z \cos \theta} \cos \nu \theta - \frac{\pi}{\sin \nu \pi}I_\nu (z)
\ee
where $I_\nu(z)$ is modified Bessel function. 
The genus one ($g= 1$) term of this series becomes
\be
U(\sigma) \sim \frac{1}{5}\sqrt{\frac{2}{3}}K_{\frac{1}{5}}(\frac{1}{2\sqrt{2}\cdot 3^{\frac{5}{2}}} \sigma^6)\sim \frac{1}{6}\sigma^{\frac{6}{5}}\Gamma(1 - \frac{1}{5}) + \cdots
\ee
which gives $\frac{1}{6}$ for the intersection numbers of the moduli space of $p=5$ spin curves.

We obtain from the equation of (\ref{p5}), the intersection numbers $<\tau_{n,j}>_g$, with condition $6(2g -1) = 5 n + j + 1$, for $p=5$,

\vskip 2mm
\noindent
(5) general $p$
\vskip 2mm
\ba
&&U(\sigma) = \frac{1}{\sigma}\oint \frac{du}{2\pi i} e^{- \sigma u^p} \nonumber\\
&\times& {\rm exp}[ - \frac{p(p-1)}{3! 4} \sigma^3 u^{p-2} - \frac{p(p-1)(p-2)(p-3)}{5!4^2}\sigma^5 u^{p-4} - \cdots]
\ea
By choosing a integral path around a cut,
\ba
U(\sigma) &=& {\rm Re} \{\frac{e^{\frac{2\pi i}{p}}}{p \sigma^{1+\frac{1}{p}}\pi} \int_0^\infty dx
x^{\frac{1}{p}-1}e^{-x} {\rm exp}[ - \frac{p(p-1)}{3! 4} \sigma^{2+ \frac{2}{p}} e^{2\pi i (1- \frac{2}{p})} x^{1-\frac{2}{p}}\nonumber\\
& -& \frac{p(p-1)(p-2)(p-3)}{5!4^2}\sigma^{4+\frac{4}{p}} e^{2\pi i ( 1-\frac{4}{p})}x^{1-\frac{4}{p}}  - \cdots]\}
\ea
We have
\ba
U(\sigma) &=& \frac{1}{\pi p \sigma^{1+\frac{1}{p}}} (\sin \frac{2\pi}{p}) \Gamma(\frac{1}{p}) + \frac{p -1}{ 24 \pi}\sigma^{1+ \frac{1}{p}}(\sin \frac{2\pi}{p}) \Gamma(1 - \frac{1}{p})\nonumber\\
&-& \frac{(p-1)(p-3)(2p+1)}{2760 \pi} \sigma^{3 + \frac{3}{p}}(\sin \frac{6\pi}{p})
\Gamma( 1 - \frac{3}{p})\nonumber\\
&-&\frac{(p-1)(p-5)(1 + 2p)(8 p^2-13p-13)}{7!4^33^2\pi}\sigma^{5+\frac{5}{p}}(\sin \frac{10 \pi}{p})\Gamma(1 - \frac{5}{p}) \nonumber\\
 &+&\frac{(p-1)(p-7)(1+2p)(72p^4-298p^3-17p^2+562p+281)}{9!4^415}\sigma^{7+\frac{7}{p}}\nonumber\\
 &\times&(\sin \frac{14\pi}{p})\Gamma(1 - \frac{7}{p})
 + \cdots
\ea
The intersection numbers of $p$ spin curves are obtained with the condition $(p+1)(2g -1) = p n + j + 1$.
\be\label{ppexpression}
U(\sigma) = \sum_g <\tau_{n,j}>_g p^{g-1} \sigma^{n+\frac{1+j}{p}}\Gamma(1 - \frac{1+j}{p}) \sin \frac {m}{p} 
\ee
with $ m = 2\pi + 4\pi (g-1)$.
\ba\label{pexpression}
&&<\tau_{1,0}>_{g=1} = \frac{p-1}{24},\hskip 3mm \nonumber\\
&& <\tau_{n,j}>_{g=2} = \frac{(p-1)(p-3)(1+2p)}{p 5! 4^2 3} \frac{\Gamma(1 - \frac{3}{p})}{\Gamma(1 - \frac{1 + j}{p})}, \nonumber\\
&&<\tau_{n,j}>_{g=3} = \frac{(p-5)(p-1)(1+ 2p)(8p^2-13p-13)}{p^2 7! 4^3 3^2}\frac{\Gamma(1- \frac{5}{p})}{\Gamma(1 - \frac{1+j}{p})}\nonumber\\
&&<\tau_{n,j}>_{g=4} = \frac{(p-1)(p-7)(1+2p)(72p^4-298p^3-17p^2+562p+281)}{p^39!4^4 15}\nonumber\\
&&\times \frac{\Gamma(1 - \frac{7}{p})}{\Gamma(1 - \frac{1+j}{p})}
\ea
This result is same as \cite{BHC3}, where the integral is restricted to a path from 0 to $\infty$
without $\sin \frac{m}{p}$ factor in $U(\sigma)$.

\vskip 2mm
\noindent
(6) $p = -1$ 
\vskip 2mm

 This expression for arbitrary $p$ in (\ref{pexpression}) allows the analytic continuation to the negative values of $p$. In the case $p= - 1$, it correspond to Euler characteristics $\chi(M_{g,1}) = \zeta(1 - 2 g)$ \cite{BHC3}. For $p= -1$, the power of $\sigma^{1+\frac{1}{p}}$ becomes zero, and $\sigma$ dependence disappears. Therefore, we need the introduction of $c$, which is taken as $N$ to specify the genus $g$.
 \ba
 U(\sigma) &=& \frac{1}{N \sigma} \int \frac{du}{2 \pi i} e^{-N {\rm log}\frac{u + \frac{1}{2}\sigma}{u - \frac{1}{2}\sigma}}\nonumber\\
 &=& \frac{1}{N}\int \frac{du}{2\pi i} (\frac{u - \frac{1}{2}}{u + \frac{1}{2}})^N\nonumber\\
 &=& \int_0^\infty dt \frac{1}{1 - e^{- t}}e^{- N t} = \sum_{n=1}^\infty (-1)^{n-1} \frac{B_n}{2n N^{2n}} 
 \ea
 We have used $\frac{u-1}{u+1}= e^{-y}$, and a following expansion,
 \be
 \frac{1}{1 - e^{-t}} = \frac{1}{t} + \frac{1}{2} + \sum_{n=1}^\infty  (-1)^{n-1} B_n\frac{t^{2n-1}}{(2n)!}
 \ee
This gives Euler characteristics (intersection number $<\tau>_g$  for $p= - 1$),
 \be
 \chi(M_{g,1}) = <\tau>_g = - \frac{1}{2g} B_g =\zeta(1 - 2g)
 \ee
 where $\zeta$ is  the Riemann zeta function and $B_n$ is a  Bernoulli number ($B_1= \frac{1}{6}$,
 $B_2 = \frac{1}{30}$, $B_3 = \frac{1}{42}$, $B_4= \frac{1}{30}$).
When $p$ is negative, we have to specify the meaning of   spin $j$. 
This index $p$ is related to the level $k$ of the Lie group $su(2)_k/u(1)$. This was studied by Witten \cite{Witten2} as a chiral ring (Landau-Ginzburg theory) of primary fields and their gravitational descendants with 
\be
p = k + 2
\ee
The present case corresponds to the singularity theory of $A_{p-1}$. When $p$ is negative,
we have  a non-compact Lorenzian group $sl(2,R)_k/u(1)$, whose discrete spectrum is
 known to correspond to two series $D_{\hat k}^{+}$ and $D_{\hat k}^{-}$ \cite{Bargmann},
The analytic continuation of $p\to - p$ corresponds to $D_{\hat k}^{-}$, and the spin $j$
takes negative value. For instance,  for $p= -1$ ,  the Euler characteristics $\chi(M_{g,1})$
is defined by the top Chern class only and $n$ , which is the power of the first Chern class, should be zero. Then we have only $<\tau_{0,-1}>_g$ in which $j=-1$.
\be\label{intersection}
<\tau_{n,j}>_g = \frac{1}{p^g}\int_{M_{g,1}} C_T(\nu) [c_1({\mathcal L})]^n
\ee
Thus $<\tau_{n,-1}>_g$ is not surprising since the discrete spectrum with negative spin exists
for $SL(2,R)$.
\vskip 2mm
\noindent
(7) $p = - 2$
\vskip 2mm

As noticed in \cite{BH7}, we have two expansions, weak coupling and strong coupling for $p= - 2$, which correspond to the Gross-Witten model for the unitary group \cite{BG,GW}. There is a phase transition between these two phases. The weak coupling corresponds to small $\sigma$ and
strong coupling corresponds to large $\sigma$. Therefore, the spin values $j$ takes negative values of $D_{\hat k}^{-}$ for weak coupling, and  positive values for the strong coupling phase. The expansion of $u(\sigma)$ is expressed by putting $n=0$ as
\be
u(\sigma) = \sum a_j \sigma^{\frac{1+j}{p}}
\ee
More details of the discrete spectrum of $SL(2,R)_k$ are presented in appendix.
Before  closing this section on GUE, we write the one point function $U(\sigma)$ as an angular
integral, which is useful for the strong coupling expansion.

In the expression of $U(\sigma)$ one puts  $\sin \theta = 1/\sqrt{1 + u^2}$ and $\cos \theta = u/\sqrt{1 + u^2}$ in (\ref{one}).
Then with $\sigma= i t$, and $u= \frac{t}{2}v$,
\be
U(\sigma) = \frac{1}{2  } \int \frac{dv}{2\pi } \exp[ - \frac{c}{p+1} (\frac{t}{2})^{p+1} \{(v+ i)^{p+1} - (v-i)^{p+1}\}]
\ee
With  $v= \frac{\cos \theta}{\sin \theta}$, it becomes 
\be
U(\sigma) = \frac{1}{2}\int_0^{\frac{\pi}{2}} \frac{d\theta}{2\pi } \frac{1}{(\sin \theta)^2} \exp [ -\frac{2 i c}{p+1}(\frac{t}{2})^{p+1}\frac{\sin(p+1)\theta}{(\sin \theta)^{p+1}}]
\ee

Note that the denominator of the exponent $(\sin \theta)^{p+1}$ becomes a numerator when
$p+1$ is negative, and it provides a large $\sigma$ expansion (strong coupling expansion for large $t$) corresponding to a discrete spectrum of $SL(2,R)_k$.
This large $t = - i \sigma$ expansion becomes, for instance for $p=-2$,
\ba
U(\sigma) &=& -\frac{1}{2}[ D - (\frac{c}{t}) + (\frac{c}{t})^2 - (\frac{c}{t})^3 + \frac{5}{6}(\frac{c}{t})^4 - \frac{7}{12}(\frac{c}{t})^5 + \cdots]\nonumber\\
&=& -\frac{1}{2} \sum_m C_m (\frac{c}{t})^m
\ea
with
\be
C_m = \frac{(2m-1)!}{m!} \frac{1}{\prod_{l=1}^{m-1} (- l^2)}
\ee
$D$ is a divergent term, which should be regularized.
The above expression matches exactly  a strong coupling expansion for the unitary (gauge) group, for a single trace result
with $N=0$ \cite{BH7}. The  unitary matrix model is
\be
Z = \int dU e^{\tr (U C^\dagger + U^\dagger C)}
\ee
where $U$ is a $N\times N$ unitary matrix, $U U^\dagger = 1$. $C$ is an external complex matrix. The strong expansion is an expansion in powers of $\tr (C^\dagger C)^m$.
The coefficient of $\tr (C^\dagger C)^m$, $C_m$ is equal to 
\ba
C_1 &=& 1, \hskip 2mm C_2 = -\frac{1}{N^2 - 1}, \hskip 2mm C_3 = \frac{4}{(N^2-1)(N^2
-4)},\nonumber\\ 
C_4 &=& -\frac{30}{(N^2-1)(N^2-4)(N^2 - 9)},...
\ea

For obtaining the $N$ dependence, we need the insertion of a logarithmic term in (\ref{one}) as  \cite{BH7},
\be
U(\sigma) = \frac{1}{2}\int \frac{du}{2 i \pi} e^{\frac{4}{\sigma (u^2-1)}}(\frac{u-1}{u+1})^N
\ee

\vskip 2mm
%%%%%%%%%%%%%%%%%%%%%%%%%%%%%%
\vskip 3mm
\section{ Classical Lie algebras }
\setcounter{equation}{0}
\renewcommand{\theequation}{4.\arabic{equation}}
\vskip 5mm
 \underline{ $X \in O(2N)$}
\vskip 2mm
When the  random matrix $X$ varies  over a classical Lie algebra, with  Gaussian distributetion, the n-point correlation function in an external source is obtained again exactly , after use of the Harish Chandra formula\cite{HC}. We have discussed in earlier work such models with external source\cite{BH4,BHL}.
   
  Consider  the Lie algebra of $O(2N)$, namely real antisymmetric matrices. Since, the Harish Chandra formula holds for this Lie algebra, we can obtain explicit expressions for the n-point correlation functions. Again one can derive a duality identity. In the present case, instead of
   the duality formula involving a  supermatrix $Q$, it is convenient to use
   \be\label{eq1}
< \prod_{\alpha=1}^k {\rm det}(\lambda_\alpha\cdot {\rm I} - X )>_A
= <\prod_{n=1}^N {\rm det}( a_n\cdot {\rm I} - Y ) >_\Lambda
\ee
where $X$ is a $2N\times 2N$ real antisymmetric matrix ($X^{t} = - X$ ) 
and $Y$ is $2k\times 2k$ real 
antisymmetric matrix ;  the eigenvalues of $X$ and $Y$ are thus pure imaginary. The  matrix source $A$ is also a $2N \times 2N$ antisymmetric matrix.  The matrix $\Lambda$ is q $2k\times 2k $ antisymmetric matrix, coupled to $Y$. We assume, without loss of  generality, that $A$ and $\Lambda$ take  the  canonical form : 

\be
A = a_1 v \oplus  \cdots \oplus a_N v, \hskip 5mm v =i\sigma_2 = \left( \matrix{0&1\cr -1&0}\right ).
\ee
$\Lambda$ is expressed also as
\be
\Lambda = \lambda_1 v \oplus \cdots \oplus \lambda_k v.\ee

The definition of the averages are
\be \langle {\cal{O}}(X) \rangle_{A} = \frac{1}{Z_{A}} \int dX \ {\cal{O}}(X)\ \exp {(\frac{1}{2}\tr X^2 + \tr A X)} \ee 
\be \langle {\cal{O}}(Y) \rangle_{\Lambda} = \frac{1}{Z_{\Lambda}} \int dY \ {\cal{O} }(Y)\ \exp {(\frac{1}{2}\tr Y^2 + \tr \Lambda Y)} \ee 

By an appropriate tuning of the  $a_n$'s, and a corresponding  rescaling of $Y$ and $\Lambda$, 
one may generate similarly  higher models of type $p$ with the conditions (\ref{condition}),
\be\label{Z_p}
Z= \int dY e^{-\frac{1}{p+1}{\tr} Y^{p+1} + {\tr} Y \Lambda}
\ee
where $p$ is an odd integer.

The HarishChandra integral for the integral over $g \in SO(2N)$ group, and given real antisymmetric matrices $Y$ and $\Lambda$,  reads  
\be\label{Th2}
\int_{SO(2N)} dg e^{{\tr}( g Y g^{-1} \Lambda)} = C
\frac{\sum\limits_{w \in W} ({\rm det}w){\rm exp}[2 \sum\limits_{j=1}^N w(y_j)\lambda_j]}{
\prod\limits_{1\leq j<k \leq N}(y_j^2-y_k^2)(\lambda_j^2-\lambda_k^2)}
\ee
where  $C=(2N-1)!\prod_{j=1}^{2N-1}(2j-1)!$, and $w$  are elements of  the Weyl group,
which consists here of permutations  followed  by reflections ($y_i\to \pm y_i\hskip 2mm ; i=1,\cdots ,N$) with an even number of sign changes.

For  the one point function, we obtain when $X$ is a $2N \times 2N$ real antisymmetric random matrix, from  the above formula,
\ba\label{U(s)}
U(\sigma) &=&  \frac{1}{2N}<{\tr} e^{\sigma X}>_A\nonumber\\
&=& \frac{1}{2N}\sum_{\alpha=1}^N \prod_{\gamma\ne \alpha}^N (\frac{(a_\alpha + \frac{\sigma}{2})^2 - a_\gamma^2}{a_\alpha^2 - a_\gamma^2}) e^{\sigma a_\alpha +\frac{\sigma^2}{4}} + (\sigma \to - \sigma)
\nonumber\\
&=&\frac{1}{N\sigma}\oint \frac{du}{2\pi i} (\frac{(u+ \frac{\sigma}{2})^2 - a_\gamma^2}{u^2 - a_\gamma^2}) \frac{u}{u+ \frac{\sigma}{4}} e^{\sigma u + \frac{\sigma^2}{4}}\nonumber\\
\ea 
where the contour encircles the poles $u= a_\gamma$.
Or,  shifting $u \to u - \frac{\sigma}{4}$,
\be
U(\sigma)= 
  \frac{1}{N\sigma} \oint \frac{dv}{2\pi i}
\prod_{i=1}^N\frac{(u - \frac{\sigma}{4})^2 - a_i^2}{(u+ \frac{\sigma}{4})^2 - a_i^2}
\left(\frac{u - \frac{\sigma}{4}}{u} \right)e^{\sigma u}
\ee
Tuning the external source to obtain the $p$-th degeneracy, one finds 
\be\label{onepoint}
U(\sigma) = \frac{1}{N\sigma} \oint \frac{du}{2i \pi} e^{-\frac{c}{p+1}[(u+\frac{\sigma}{4})^{p+1}
- (u -\frac{\sigma}{4})^{p+1}]} (1 - \frac{\sigma}{4 u})
\ee
\vskip 3mm
\noindent
(1) $p=3$
\vskip 2mm
 There are two terms in (\ref{onepoint}) ; the first term $U(\sigma)^{OR}$ is exactly one-half
 of $U(\frac{\sigma}{2})$ for the
GUE (orientable Riemann surfaces). The second term is a new term, and we denote it as the non-orientable part U($\sigma$)$^{NO}$, since it is related to non-orientable surfaces with half-integer genus : 

\ba
U(\sigma)^{OR} &=& \frac{1}{2}U(\frac{\sigma}{2})\nonumber\\
&=&  \frac{1}{12\sqrt{3}}[ J_{\frac{1}{3}}(\frac{1}{12\sqrt{3}}(\frac{\sigma}{2})^4) + J_{-\frac{1}{3}}(\frac{1}{12\sqrt{3}}(\frac{\sigma}{2})^4)]\nonumber\\
 &=& \frac{1}{2\cdot3^{\frac{1}{3}}(\frac{\sigma}{2})^{\frac{4}{3}}}Ai( - \frac{1}{4\cdot 3^{1/3}}(\frac{\sigma}{2})^{\frac{8}{3}})
\ea

For the non-orientable surfaces, from the condition,
\be
(p+1)(2g - 1) = p n + j +1
\ee
we find that the genus $g$ is always a half-integer ($g= \frac{1}{2},\frac{3}{2},\frac{5}{2},...$), and U($\sigma$)$^{NO}$ has a series expansion in powers   of $\sigma^{n + \frac{1+j}{p}}$. For $p=3$, we have
\ba
U(\sigma)^{NO} &=& \frac{1}{4}\oint \frac{du}{2\pi i}\frac{1}{u} e^{- \frac{\sigma }{2} u^3 - \frac{\sigma^3}{32}u}\nonumber\\
&=& \frac{1}{12}\oint \frac{dx}{2\pi i} \frac{1}{x}e^{- x - \frac{2^{1/3}}{32} \sigma^{8/3} x^{1/3}}\nonumber\\
&=& {\rm Re} \{ \frac{1}{12 i \pi}\int_0^\infty \frac{1}{x} e^{- x - \frac{1}{4}e^{\frac{2\pi i}{3}}(\frac{\sigma}{2})^{8/3} x^{1/3}} \}
\ea
This function may be expanded as
\ba
U(\sigma)^{NO} &=& {\rm Re} \{ \frac{1}{12 i \pi} \int_0^\infty dx \frac{1}{x}  e^{-x}
\sum_{n=0}^\infty \frac{1}{n!} (-\frac{1}{4} e^{\frac{2\pi i}{3}}(\frac{\sigma}{2})^{8/3}x^{1/3})^n \}\nonumber\\
&=& -\frac{1}{\pi} \frac{1}{48}(\frac{\sigma}{2})^{\frac{8}{3}}(\sin \frac{2\pi}{3})\Gamma(\frac{1}{3})\nonumber\\
&+& \frac{1}{\pi} \frac{1}{384} (\frac{\sigma}{2})^{\frac{16}{3}}(\sin \frac{4\pi}{3})
\Gamma(\frac{2}{3}) - \frac{1}{\pi}\frac{1}{3! \cdot 12\cdot4^3}(\frac{\sigma}{2})^{8}
(\sin 2\pi) \nonumber\\
&+& \frac{1}{\pi} \frac{1}{4!\cdot 12\cdot 4^4}(\frac{\sigma}{2})^{\frac{32}{3}}(\sin \frac{8\pi}{3})\Gamma(\frac{4}{3}) - \cdots
\ea

Using Airy functions, the  $p=3$ case is expressed as
 \be\label{IntAiry}
 U(\sigma) = \frac{1}{2\cdot3^{1/3}(\frac{\sigma}{2})^{4/3}}Ai(x) - \frac{1}{4}\int_0^x dx^\prime Ai(x^\prime)
 \ee
 with $ x = -\frac{1}{4\cdot 3^{1/3}} (\frac{\sigma}{2})^{8/3}$.
  \vskip 2mm

The Airy function $Ai(z)$ and the integral of Airy function  may be  expanded as
\be
Ai(z) = \frac{\pi}{3^{2/3}}\sum_{n=0}^\infty \frac{1}{n! \Gamma(n + \frac{2}{3})}(\frac{1}{3})^{2n} z^{3n} - \frac{\pi}{3^{4/3}} \sum_0^\infty \frac{1}{n! \Gamma(n + \frac{4}{3})}
(\frac{1}{3})^{2n} z^{3n+1}
\ee
\be
\int_0^z Ai(t) dt = \frac{\pi}{3^{2/3} \Gamma(\frac{2}{3})}z - \frac{\pi}{3^{4/3}\cdot 2
\Gamma(\frac{4}{3})} z^2 + \frac{\pi}{36\cdot 3^{2/3} \Gamma(\frac{5}{3})} z^4 + \cdots
\ee

Inserting these expansions, we have for $p=3$, $ \frac{\pi}{\sin (\frac{\pi}{3})}= \Gamma(\frac{1}{3})\Gamma(\frac{2}{3})$.
\ba
U(\sigma) &=& 
\frac{\pi}{24\Gamma(\frac{1}{3})} (\frac{\sigma}{2})^{4/3} - \frac{\pi}{108\cdot 64 \Gamma(\frac{2}{3})} (\frac{\sigma}{2})^{20/3} + \cdots\nonumber\\
&+& [ \frac{\pi}{48 \Gamma(\frac{2}{3})} (\frac{\sigma}{2})^{8/3} + \frac{\pi}{384 \Gamma(\frac{1}{3})}(\frac{\sigma}{2})^{16/3} - \frac{\pi}{864\cdot 4^4 \Gamma(\frac{2}{3})} (\frac{\sigma}{2})^{32/3} + \cdots ] \nonumber\\
\ea
\ba
U(\sigma) &=& <\tau_{1,0}>_{g=1} \Gamma(1 - \frac{1}{3}) (\frac{\sigma}{2})^{1 + \frac{1}{3}} + 
<\tau_{6,1}>_{g=3}\Gamma(1-\frac{2}{3})  3^2(\frac{\sigma}{2})^{6+\frac{2}{3}} + \cdots\nonumber\\
&+& [\hskip 1mm 
<\tau_{2,1}>_{g=3/2} \Gamma( 1 - \frac{2}{3}) 3^2 (\frac{\sigma}{2})^{2 + \frac{2}{3}} + <\tau_{5,0}>_{g=5/2} \Gamma(1 - \frac{1}{3}) 3^4 (\frac{\sigma}{2})^{16/3} \nonumber\\
&&+ <\tau_{10,1}>_{g=9/2} \Gamma(1 - \frac{2}{3}) 3^8 (\frac{\sigma}{2})^{32/3} + \cdots]
\ea
 We have for $p=3$,
 \ba\label{exp1}
 U(\sigma)^{NO} &=& \frac{1}{12}  y^2\Gamma(1-\frac{2}{3})  + \frac{1}{24} y^4  \Gamma(1 - \frac{1}{3}) \nonumber\\
&&+ \frac{1}{864} y^8 \Gamma(1 - \frac{2}{3})  + ....
\ea

We have obtained for $p=3$ the explicit intersection numbers for non-orientable surfaces  with  one marked point.
The intersection number $<\tau_{2,1}>_{g=3/2}$ corresponds to a cross-capped torus.
For $g=1/2$
wa are dealing with the topology of the projective plane  but for  this case, the intersection numbers  $<\tau_{0,1}^2>_{g=1/2}$ are present only beyond the two marked points level \cite{BH4}.
We have
\be
<\tau_{1,0}>_{g=1} = \frac{1}{24}, \hskip 2mm <\tau_{2,1}>_{g=\frac{3}{2}}= \frac{1}{864}, \hskip 2mm...
\ee
\vskip 2mm

\vskip 3mm
\noindent
(2) general $p$
\vskip 2mm

 Using the binomial expansion, one finds ($y = 2^{\frac{1}{p}}(\frac{\sigma}{4})^{1+ \frac{1}{p}} = \frac{1}{2}(\frac{\sigma}{2})^{1+\frac{1}{p}}$)
\ba
U(\sigma) = &&-\frac{1}{4 y p N} \int dt t^{\frac{1}{p} - 1} e^{-t} 
[ 1 - \frac{p(p-1)}{6}y^2 t^{1 - \frac{2}{p}} + \cdots ]\nonumber\\
&& \times [ 1 + y t^{-\frac{1}{p}}]
\ea

This is again the sum of  two contributions, orientable (OR) and non-orientable (NO). The odd powers in $y$ correspond to the orientable contribution, which is the same as for the unitary case ; 
 the even powers in $y$ correspond to the non-orientable case :
\be\label{O2N}
U(\sigma) = U(\sigma)^{OR} + U(\sigma)^{NO}
\ee
$U(\sigma)^{OR}$ is same as GUE but the normalization of $\sigma$ is replaced by $\sigma/2$.

The first term in the above series expansion is divergent, and it should be regularized.
Except for this divergent term, we give the series expansion  up to order  $y^8$ (we have neglected the phase factor $\sin(\frac{2\pi m}{p})$,
\ba\label{UUNO}
U(\sigma)^{NO} &= &\frac{y^2}{24} (p-1) \Gamma(1 - \frac{2}{p})\nonumber\\
&+& \frac{y^4}{6!}(p-1) (p^2 - 5 p + 1) \Gamma(1 -\frac{4}{p})\nonumber\\
&+& \frac{y^6}{7!\cdot 9}(p-1)(p-3)(4 p^3 -23 p^2 - 2p - 6) \Gamma(1 - \frac{6}{p})\nonumber\\
&+& \frac{y^8}{7! 3^3 \cdot 10} (p-1)(9 p^6 - 121 p^5 + 435 p^4 - 317 p^3\nonumber\\
& & - 167 p^2 - 471p - 43)\Gamma( 1 - \frac{8}{p}) + O(y^{10})
\ea
From this genus expansion, one obtains the intersection numbers of $p$-spin curves for  non-orientable surfaces.
\vskip 3mm
\noindent
(3) $p = - 1$
\vskip 3mm
We now perform the limit, $p\to -1$, which is related to the virtual Euler characteristics.
When we put $p= -1$ in (\ref{UUNO}),   the $\Gamma$ function term becomes  an integer for $p=-1$, and this agrees with the intersection number of $<\tau_{1,0}>_g$, which gives a factor $\Gamma(1 - \frac{1}{p}) = \Gamma(2)=1$ for the spin zero. We obtain
\ba\label{y2g}
U(\sigma)^{NO} &=& - \frac{1}{24} (2y)^2 
 - \frac{7}{240} (2y)^4
 - \frac{31}{504} (2y)^6
- \frac{127}{480} (2 y)^8
+ \cdots
\ea
This series agrees precisely with  the   series expansion 
\be
U(\sigma)^{NO} = - \sum_{\hat g=1}^\infty \frac{1}{2\hat g} (2^{2\hat g - 2} - \frac{1}{2}) B_{\hat g} (2 y)^{2\hat g}
\ee
where $B_{\hat g}$ is  a Bernoulli number, a  positive rational number. $B_1 = \frac{1}{6}, B_2 = \frac{1}{30}, B_3=\frac{1}{42}, B_4 =  \frac{1}{30}$.
The coefficient of $(2 y)^{2\hat g}$ is the same as for the virtual  Euler characteristics of the moduli space of real algebraic curves for  genus $g$ and  one marked point, which was derived from the Penner model of the real symmetric matrix by Goulden et al. \cite{Goulden}. ( We use  for the half genuses in the list , $\frac{1}{2}$, $1$, $\frac{3}{2}$, $2$,... for a projective plane, Klein bottle, cross-capped torus, doubly cross-capped torus ,..., with the notation $\hat g=1$, $\hat  g=2$, $\hat g=3$,$\hat g=4$,....,respectively \cite{Jackson}, and this is a reason for  the appearance of the $(2y)^{2\hat g}$ factor in (\ref{y2g})).

Since we derived this from
the antisymmetric $O(2N)$ Lie algebra, the coincidence between $O(2N)$ lie algebra and GOE for the virtual Euler characteristics seems remarkable.
\be\label{NO1}
\chi^{NO}(\bar M_{g,1}) =  \frac{1}{2g}(\frac{1}{2} - 2^{2g-2}) B_{g}.
\ee
This result may be  obtained analytically to all orders. We now derive this result  
 from the  integral form (\ref{onepoint})  replacing $c$ by $N$. With $p=-1$, it becomes
\be
U(\sigma) = - \frac{1}{4 N \sigma} \int du (\frac{u - \sigma}{u + \sigma})^N ( 1 + \frac{\sigma}{ u})
\ee
With the change of variable $u \to \sigma u$,
\be
U(s) = -\frac{1}{4 N} \int du (\frac{u -1}{u + 1})^N ( 1 + \frac{1}{u})
\ee
We divide it into two parts, $U(\sigma)^{OR}$ and $U(\sigma)^{NO}$,
\be
U(\sigma)^{OR} = - \frac{1}{4N}\int du (\frac{u-1}{u+1})^N
\ee
\be
U(\sigma)^{NO} = - \frac{1}{4N} \int du (\frac{u-1}{u+1})^N \frac{1}{u}
\ee
We use the same change of variables as for the unitary case \cite{BH3},
\be
\frac{u-1}{u+1} = e^{- y},\hskip 5mm u= \frac{1 + e^{-y}}{1 - e^{- y}}, \hskip 5mm
du = -2 \frac{e^{-y}}{(1 - e^{-y})^2} dy
\ee

\be
U(\sigma)^{OR} = \frac{1}{2N} \int dy e^{-Ny} \frac{e^{-y}}{(1 - e^{-y})^2} 
\ee
\ba\label{both}
U(\sigma)^{NO} &=& \frac{1}{2N}\int dy e^{-Ny} \frac{e^{-y}}{(1 - e^{-y})^2} (\frac{1 - e^{-y}}{1 + e^{-y}})\nonumber\\
&=& \frac{1}{4N} \int dy e^{-Ny} [ \frac{1}{1 - e^{-y}} - \frac{1}{1 + e^{-y}} ]
\ea

It is  interesting to note that both  Boson and Fermion distributions  enter
in the above integrand (\ref{both}).

If we  use the expansions,
\ba\label{Bernoulli}
\frac{1}{1 - e^{-y}} &=& \frac{1}{y} + \frac{1}{2}+ \sum_{n=1}^\infty (-1)^{n-1}\frac{ B_{n}}{(2n)!} y^{2n-1}\nonumber\\
\frac{1}{1 + e^{- y}} &=& \frac{1}{2} +  \sum_{n=1}^\infty \frac{(-1)^{n-1}(2^{2n} - 1)}{(2n)!}  B_n y^{2n -1}
\ea
then they become
\ba
U(\sigma)^{OR} &=& \frac{1}{2N}\int dy \frac{1}{y^2} e^{- N y}
      - \frac{1}{2} \sum_{n=1}^\infty (-1)^{n-1} \frac{B_n}{2n} \frac{1}{N^{2n}}\nonumber\\
U(\sigma)^{NO} &=& \frac{1}{4N}\int dy e^{-Ny}  + \frac{1}{4}\sum_{n=1}^\infty (-1)^{n-1} \frac{B_n}{2n} \frac{1}{N^{2n+1}}\nonumber\\
&-& \frac{1}{4}\sum_{n=1}^\infty (-1)^{n-1}\frac{(2^{2n}-1)}{2n} B_n \frac{1}{N^{2n+1}}\nonumber\\
&=& \frac{1}{4N}\int dy \frac{e^{-Ny}}{y} + \frac{1}{4}\sum_{n=1}^\infty (-1)^{n-1} \frac{(2 - 2^{2n}) B_n}{2n}\frac{1}{N^{2n+1}}
\ea

We now get from the above equation (replacing $n$ by $g$),
\ba\label{chi}
\chi^{OR}(\bar M_{g,1}) &=& - \frac{1}{2}\zeta(1 - 2g) =  - \frac{1}{2}\frac{(-1)^g B_{g}}{2g},\nonumber\\
\chi^{NO}(\bar M_{ g,1}) &=& (-1)^{g-1}\frac{1}{2 g}(2^{2 g-2} - 2^{-1}) B_{ g}
\ea

For s marked point, the result obtained from the real symmetric matrix Penner model \cite{Goulden} is
\be\label{EulerNOP}
\chi^{NO}(\bar M_{g,s}) = (-1)^s\frac{1}{2}\frac{(2g + s - 2)! (2^{2g-1} - 1)}{(2g)!s!} B_{g}
\ee
 This result can be obtained by applying  equation (\ref{chi}) \cite{BH5}. 
In this O(2N) model, we have  the following condition,  the same as for Riemann surfaces with spin $j$ and  s-marked points 
\be
(p+1)(2 g - 2 + s) = p \sum_{i=1}^s n_i + \sum_{i=1}^s  j_i + s
\ee
However, we have to assign the genus $g$ also to half integers  to represent non-orientable surfaces \cite{BH4}.

\vskip 2mm
\noindent
 \underline{ $ X \in O(2N+1)$}
\vskip 3mm
For $SO(2N+1)$ Lie algebra, the matrix $X$ is
\be
X = h_1 v \oplus h_2 v \oplus \cdots h_N v \oplus 0
\ee
The measure is $V(H)^2$,
\be
V(H) = \prod_{1\leq j\leq N}(h_j^2 - h_k^2) \prod_{j=1}^N h_j
\ee
The Harish Chandra formula is
\be
I = \int_{SO(2N+1)}e^{{\rm tr} (g a g^{-1} b )} dg = C_{G(N)} \frac{\sum\limits_{w\in G(N)}({\rm det} w) {\rm exp}( 2 \sum\limits_{j=1}^N w(a_j)b_j)}{\prod\limits_{1\leq j \leq k\leq N}(a_j^2-a_k^2)(b_j^2 - b_k^2)\prod\limits_{j=1}^N a_j b_j}
\ee
with $C_{G(n)} = \prod_{j=1}^N (2j -1)! \prod_{j=2N}^{4N-1}j!$.
Comparing with the  $O(2N)$ case, this formula differs from (\ref{Th2}) by the presence of the term
$\prod a_j b_j$  in the denominator.
For the one point function, we have
\be
U(\sigma) = \frac{1}{N}\sum_{\alpha=1}^N \int_{-\infty}^\infty \prod_{i=1}^N d\lambda_i
\frac{\prod (\lambda_i^2 - \lambda_j^2)\prod \lambda_k}{\prod (a_i^2 - a_j^2)\prod a_k}
e^{- \sum \lambda_i^2 + \sigma \lambda_\alpha + 2 \sum a_i \lambda_i}
\ee
This sum of integrals may be written as a contour integral, which collects poles at $u= a_i^2$,
\ba
U(\sigma) &=& \oint_{\{u= a_i^2\}} \frac{du}{2\pi i} \prod_{j=1}^N \frac{(\sqrt{u}+\sigma)^2 - a_j^2}{u - a_j^2} \frac{1}{(\sqrt{u}+\sigma)^2 - u}(1 + \frac{\sigma}{\sqrt{u}}) e^{\sigma^2 + 2 \sigma \sqrt{u}}\nonumber\\
&=&\frac{2}{\sigma}\oint \frac{dv}{2\pi i} \prod_{j=1}^N \frac{(v+ \sigma)^2 - a_j^2}{v^2 - a_j^2} \frac{v + \sigma}{\sigma + 2 v} e^ {\sigma^2 + 2\sigma v}\nonumber\\
&=& \frac{1}{\sigma} \oint \frac{dv}{2\pi i} \prod_{j=1}^N \frac{(v + \frac{\sigma}{2})^2 - a_j^2}{(v - \frac{\sigma}{2})^2 - a_j^2} (1 + \frac{\sigma}{2 v}) e^{\sigma v}
\ea
By the tuning  to the $p$-th degeneracy, we obtain
\be
U(\sigma) = \frac{1}{\sigma}\oint \frac{du}{2\pi i} e^{- \frac{1}{p+1}((u + \frac{\sigma}{2})^{p+1} - (u - \frac{\sigma}{2})^{p+1})}( 1 + \frac{\sigma}{2 u})
\ee
This takes the  same form as for the $O(2N)$ case.

\vskip 3mm

\vskip 3mm
\noindent
\underline{ $ X \in Sp(N)$}
\vskip 3mm
The Haar measure of $Sp(N)$ is $\Delta(\lambda)^2$, with
\be
\Delta(\lambda) = \prod_{i<j}(\lambda_i^2 - \lambda_j^2) \prod_k \lambda_k
\ee

The Harish Chandra formula for $Sp(N)$ reads \cite{BHL}
\ba
I &=& \int_G e^{<Ad(g)\cdot a|b>}dg  = \frac{\sum_{w\in W}({\rm det} w) e^{<w\cdot a|b>}}{\Delta(a)\Delta(b)}\nonumber\\
&=&C \frac{{\rm det}[ 2 {\rm sinh}(2 a_i b_j)]}{\prod (a_i^2 - a_j^2)(b_i^2 - b_j^2) \prod (a_k b_k)}
\ea

For the  one point function, we have
\ba
U(\sigma) &=&\frac{1}{N}\sum\limits_{\alpha=1}^N \int_{-\infty}^\infty \prod_{i=1}^N d\lambda_i \frac{\prod\limits_{1\leq i < j\leq N}(\lambda_i^2 -\lambda_j^2)\prod\limits_{1\leq k\leq N} \lambda_k}{\prod\limits_{1\leq i < j\leq N}(a_i^2 - a_j^2) \prod\limits_{1\leq k\leq N} a_k} e^{-\sum \lambda_i^2 + \sigma \lambda_\alpha + 2\sum a_i \lambda_i}\nonumber\\
&=&\oint \frac{du}{2\pi i}\prod_{j=1}^N \frac{(\sqrt{u} +\sigma)^2 - a_j^2}{u - a_j^2} \frac{1}{(\sqrt{u} + \sigma)^2 - u} (1 + \frac{\sigma}{\sqrt{u}}) e^{\sigma^2 + 2 \sigma \sqrt{u}}\nonumber\\
&=&\frac{2}{\sigma}\oint \frac{dv}{2\pi i} \prod_{j=1}^N \frac{(v+ \sigma)^2 - a_j^2}{v^2 - a_j^2} \frac{v + \sigma}{\sigma + 2 v} e^ {\sigma^2 + 2\sigma v}\nonumber\\
&=& \frac{1}{\sigma} \oint \frac{dv}{2\pi i} \prod_{j=1}^N \frac{(v + \frac{\sigma}{2})^2 - a_j^2}{(v - \frac{\sigma}{2})^2 - a_j^2} (1 + \frac{\sigma}{2 v}) e^{\sigma v}
\ea
where we have shifted  $v\to v-\frac{\sigma}{2}$ and $a_\gamma \to a_\gamma/2$. This expression becomes the same as for the  $O(2N)$ case,  when we put $v \to 2 v$ up to  a factor 2. Note that we do not need to consider the  expansion $\frac{\sigma}{2}$ as  in the $O(2N)$ case. The first term of the expression is  same as for GUE. By the tuning $a_\gamma$ to the $p$-th case, we have
\be
U(\sigma) = \frac{1}{\sigma}\oint \frac{du}{2\pi i} e^{- \frac{1}{p+1}((u + \frac{\sigma}{2})^{p+1} - (u - \frac{\sigma}{2})^{p+1})}( 1 + \frac{\sigma}{2 u})
\ee
We write these two terms as $U(\sigma) = U(\sigma)^{OR} + U(\sigma)^{NO}$. It is then obvious that we obtain the same intersection numbers and virtual Euler characteristics  as in the $O(2N)$ case.

%%%%%%%%%%%%%%%%%%%%%%%%%%%%%%%%%%%%%%%%%%
\section{ Open intersection numbers}
\setcounter{equation}{0}
\renewcommand{\theequation}{5.\arabic{equation}}
\vskip 5mm
\noindent{\bf{Kontsevich-Penner model}}
\vskip 3mm
 The Airy matrix model with an external source, the Kontsevich model for $p=2$, gives the intersection numbers for  closed Riemann surfaces, which satisfy a  KdV hierarchy. These closed intersection numbers are obtained from (\ref{ppexpression}) and (\ref{pexpression}) for one marked point. They are  known for genus $g$ and one marked point in a simple closed form, 
 \be\label{Kontsevich}
 <\tau_{3g-2,0}>_g = \frac{1}{(24)^g g!}
 \ee
When the Riemann surface has boundaries, open intersection numbers appear, which differ from  that of the Kontsevich model. We have  studied earlier the effect of an additional logarithmic potential in the  Kontsevich model, the so called Kontsevich-Penner model \cite{BH07}. In our work this model came  from a two matrix model,  which originated itself  from a time-dependent matrix model. The eigenvalues of  the two matrices correspond for one  to  the edge of the distribution and for the other one to the bulk. Then we can use the duality identity for  the two characteristic polynomials of the two matrices  with external sources,  and thereby recover the Kontsevich-Penner model. Therefore the presence in that model of the term i of $({\rm det} M)^k = {\rm exp}[k \tr {\rm log} M]$ coreesponds to the addition of   a boundary (an open disc) in the random surfaces  described by the Kontsevich model.  
Recently the open intersection numbers have been  analyzed in 
\cite{Pandharipande, Buryak1,Buryak2}.  The generating matrix model for those open intersection numbers are given by a  Kontsevich-Penner model\cite{Alexandrov1,Alexandrov2}. This Kontsevich-Penner model has different Virasoro equations and different intersection numbers, which depend upon an additional  parameter $k$ which corresponds to the logarithmic term 

\be\label{KontsevichPenner}
Z 
=\int dM e^{\frac{1}{3}\tr M^3 + \tr M\Lambda + k \tr {\rm log} M}
\ee
For the open intersection numbers, considered by \cite{Pandharipande}, $k$ takes the value  $k=1$ \cite{Alexandrov2}.
The addition of the logarithmic potential yields  new Virasoro equations and new intersection numbers which related to the boundary insertions. The intersection numbers  for the model (\ref{KontsevichPenner}) have been computed in   \cite{BH07},
\be\label{tau1}
<\tau_1>_{g=1} = \frac{1+12 k^2}{24},
\hskip 3mm <\tau_0\tau_{\frac{1}{2}}>_{g=\frac{1}{2}} = k, \hskip 3mm ...
\ee
The appearance of a half-integer index exhibits the non-orientable nature. The non-vanishing $<\tau_{n_1}\cdots \tau_{n_s}>_{g}$ are restricted  by the condition
\be
3 (2g - 2+ s) = 2\sum_{i=1}^s n_i + s
\ee
When the parameter $k$ vanshes, the intersection numbers reduce to  the usual Kontsevich result, which satisfies a KdV hierarchy. When the cubic Airy matrix part is absent, and only the logarithmic potential  is present (Penner model), as we have seen in the $p=-1$ case in section 3, the model gives the  Euler characterstics \cite{BHC3}. When $k=1$, it reduces to open intersection numbers. The meaning of the parameter $k$ is found in  the two matrix model \cite{BHC3, BH07}.

We now consider the $k$-dependence with one marked point. The one point intersection numbers of 
the Kontsevich-Penner model  (\ref{KontsevichPenner}) are  obtained from  $U(\sigma)$  \cite{BH07}
\ba\label{KP}
U(\sigma) &=& \frac{1}{\sigma} \oint \frac{du}{2\pi i} e^{- \frac{c}{3}[(u+ \frac{\sigma}{2})^3 - (u - \frac{\sigma}{2})^3] + k {\rm log}(u+ \frac{\sigma}{2}) - k {\rm log}(u - \frac{\sigma}{2})}\nonumber\\
&=& \frac{1}{ \sigma} e^{-\frac{ c}{12}\sigma^3}\oint \frac{du}{2\pi i}e^{-{c \sigma} u^2 + k {\rm log}(u + \frac{1}{2}\sigma) - k {\rm log}(u-\frac{1}{2}\sigma)}
\ea
with
\be
\sigma = \frac{1}{\lambda},\hskip 3mm t_n = \frac{1}{\lambda^{n + \frac{1}{2}}}
\ee
This $U(\sigma)$ correctly reduces to the intersection numbers of  the Kontsevich model
with one marked point when $k=0$,
\be
U(\sigma) = \frac{\sqrt{\pi}}{\sqrt{c}} \sum_{g=1}^\infty \frac{(-c)^g}{(12)^g g!}t_{3g-2}
\ee
Including the factor $\frac{1}{(-cp)^{g}}$, ($p$=2), the intersection number reduces to 
\be
<\tau_{3g-2}>_g = \frac{1}{(24)^g g!}
\ee

For dealing with higher $k$'s, we expand (\ref{KP}), after rescaling of $u$,

\ba\label{residue}
U(\sigma) &=& \frac{1}{2 \sigma^{\frac{3}{2}}}e^{ - \frac{c \sigma^3}{12}}\oint \frac{du}{2\pi i}
e^{- \frac{c}{4}u^2}[ 1 +  k (\frac{2 }{u} \sigma^{\frac{3}{2}} + \frac{2}{3 u^3}\sigma^{\frac{9}{2}} + \frac{2}{5 u^5}\sigma^{\frac{15}{2}} \cdots)\nonumber\\
&+& k^2 (\frac{2}{u^2}\sigma^3 + \frac{4 }{3 u^4}\sigma^6 + \cdots)\nonumber\\
&+& k^3 (\frac {8}{3! u^3}\sigma^{\frac{9}{2}} + \cdots) + k^4 (\frac{16 }{4! u^4} \sigma^6 +\cdots) + O(k^5)]
\ea

The coefficients of the successive orders in  $k$ tmay be computed from 
\ba
&&k\frac{ e^{-\frac{c}{12}\sigma^3}}{\sigma^{\frac{3}{2}}}\oint \frac{du}{2\pi i}e^{- \frac{c}{4} u^2} {\rm log}
\frac{u+\sigma^{3/2}}{u-\sigma^{3/2}}\nonumber\\
&&= k\frac{1}{\sigma^{\frac{3}{2}}}\sqrt{\frac{\pi}{c}} e^{-\frac{c}{12}\sigma^3}
{\rm erf} (\frac{\sqrt{c}}{2}\sigma^{3/2})
\ea
with
\be
{\rm erf}(x) = \frac{2}{\sqrt{\pi}}\int_0^x e^{- t^2} dt
\ee
where the integral is computed  as the discontinuity across the   cut between -1 to 1 in  the $u$-plane. This integral becomes a contour integral around $u=0$ by expanding the logarithm in powers of $\frac{1}{u}$ as
\ba\label{oddpower}
&&\frac{k}{\sigma^{3/2}}e^{-\frac{c \sigma^3}{12}}\oint \frac{du}{2\pi i}e^{-\frac{c}{4} u^2}
 \sum_{n=0}^\infty \frac{\sigma^{\frac{3}{2}(2n +1)}}{ (2n+1)u^{2n+1}}\nonumber\\
&&= k e^{-\frac{c}{12}\sigma^3}\sum_{n=0}^\infty
 \frac{1}{n! (2n+1)} (-\frac{c \sigma^3}{4})^n
\ea
For the odd powers of $k$, the integration over $u$ is the same contour integral around 
$u=0$ \cite{BH07}.

 For the even powers of $k$, we use the following integrals,
\be\label{even}
\int_{-\infty}^\infty du e^{- a u^2} \frac{1}{u^{2n}}= (-1)^n\frac{2^n \sqrt{\pi}}{(2n-1)!!} a^{\frac{2n-1}{2}}
\ee
which may be obtained by  integration over $a$.
Putting $a=\frac{c}{4}$ we obtain
\be
\int e^{-\frac{c}{4} u^2} \frac{1}{u^{2}} du = -\sqrt{c\pi}
\ee
Thus we obtain, up to  terms of order $k^2 \sigma^3$,
\ba
U(\sigma) &=& e^{-\frac{c \sigma^3}{12}}  \frac{1}{2\sigma^{\frac{3}{2}}}
\int \frac{du}{2\pi } e^{-\frac{c}{4}u^2}[ 1 +  2 k^2 \sigma^3\frac{1}{u^2} ]
\nonumber\\
&=& \frac{1}{2\pi}\sqrt{\frac{\pi}{c}}e^{-\frac{c}{12}\sigma^3}\frac{1}{\sigma^{\frac{3}{2}}}(1 - c k^2 \sigma^3)\nonumber\\
\ea
Expanding the factor $e^{-\frac{c}{12}\sigma^3}$, we obtain the intersection number 
$<\tau_1>$ as
\be
<\tau_1>_{g=1} = \frac{1}{24} (1 + 12 k^2)
\ee
For $<\tau_4>_{g=2}$ and $<\tau_{7}>_{g=3}$ , we obtain with (\ref{even})
\ba\label{tau4}
&&<\tau_4>_{g=2}= \frac{1}{1152}(1 + 56 k^2 + 16 k^4), \nonumber\\
&&<\tau_{7}>_{g=3} = \frac{1}{2073600}(25 + 5508 k^2 + 3120 k^4 + 192 k^6)
\ea

The intersection numbers for fractional genus, $< \tau_{\frac{5}{2}} >_{g=\frac{3}{2}}$, $<\tau_{\frac{11}{2}}>_{g=\frac{5}{2}}$, ... are expressed as 
polynomials with  odd powers of $k$ and they are given  by the residues for the terms of order $\sigma^3$, $\sigma^6$,... in
(\ref{residue}).
\be
< \tau_{\frac{5}{2}} >_{g=\frac{3}{2}} = \frac{1}{12} ( k + k^3)
\ee
In \cite{BH07}, there is  a misprint for this term of order $\sigma^{3/2}$, which had been  evaluated from Virasoro equations. The above results agree with the Virasoro equations,  which  will be discussed below. 
In general, the intersection numbers with one marked point
$<\tau_{3g -2}>_{g}$ are easily computed to all orders  by using the formulae  (\ref{oddpower}) and (\ref{even}).

We have used the condition corresponding   to $p=2$ and one point, $s=1$, $(p+1)(2g -1) = 2n +1$ for $<\tau_n>_g$. This condition $3(2g-1)= 2n +1$ implies $n=3g -2$ and if $n$ is a half integer, then the genus $g$ is also  a half integer. Those half integer $g$ appear for non-orientable surfaces, as discussed
earlier with the  random surfaces generated by antisymmetric matrices \cite{BH4} in section 5, and it corresponds to the   topology of non orientable surfaces
such as  the  projective plane ($g=\frac{1}{2}$), the Klein bottle ($g=1$), the cross-capped torus ($g=\frac{3}{2}$), etc. \cite{Jackson}.

 The string equation for the Kontsevich-Penner model has been derived in \cite{BH07,Alexandrov2}; it reads
 \ba\label{string1}
\frac{\partial F}{\partial t_0} &=& \sum_{n=0,1,2,...} (n + \frac{1}{2}) t_{n+1}\frac{\partial F}{\partial t_n} + \sum_{n=\frac{1}{2},\frac{3}{2},...} (n + \frac{1}{2}) t_{n+1}\frac{\partial F}{\partial t_n} \nonumber\\
&+& \frac{1}{4} t_0^2 - \frac{k}{2}t_{\frac{1}{2}}
\ea
The free energy $F$ is divided into close and open parts, $F^c$ and $F^o$.  Then, we have
\ba\label{string1b}
&&\frac{\partial F^c}{\partial t_0} = \frac{1}{4} t_0^2 + \sum_{n=0,1,2,..} (n + \frac{1}{2})t_{n+1}\frac{\partial F}{\partial t_n},\nonumber\\
&&\frac{\partial F^o}{\partial t_0} =  - \frac{k}{2}t_{\frac{1}{2}}  + \sum_{n=\frac{1}{2},\frac{3}{2},...} (n + \frac{1}{2})t_{n+1}\frac{\partial F}{\partial t_n}
\ea
The Virasoro equations for open intersection theory for genus zero has been discussed in \cite{Pandharipande,Buryak1}.
The open intersection numbers are defined  analogously to the closed case as 
\be\label{string2}
<\tau_{n_1} \tau_{n_2} \cdots \tau_{n_s} \hat\sigma^{\hat k}>_{g}^o
= \int_{\overline M_{g,{\hat k},s}} \psi_1^{n_1}\psi_2^{n_2} \cdots \psi_{s}^{n_s}
\ee
The string equation becomes for the open free energy is
\be
\frac{\partial F^o}{\partial \hat t_0} = \sum_{i=0}^\infty \hat t_{i+1}\frac{\partial F^o}{\partial \hat t_i} + \hat s
\ee
which is consistent with (\ref{string1b}) ($\hat s$ is proportional to $k$, and the difference is due to a different normalization of $\hat t_n$ ). The string equation implies
\be\label{string3}
< \tau_0 \prod \tau_{n_i} \hat \sigma^{\hat k}>_g^o = \sum_j <\tau_{n_j-1} \prod_{i\ne j}
\tau_{n_i}\hat \sigma^{\hat k}>_g^o
\ee
 We will consider the case of  two marked points and derive this string equation in the next section.
 
\vskip 3mm
\noindent{\bf{open $p$-th spin cuves}}
\vskip 3mm

The  open intersection numbers for the $p$-th spin curves with  boundaries are also  given by the
addition of  a logarithmic potential to (\ref{one})
\be
U(\sigma) = \frac{1}{\sigma} \oint \frac{du}{2\pi} e^{-\frac{c}{p+1}[(u + \frac{\sigma}{2})^{p+1} - (u - \frac{\sigma}{2})^{p+1}]  + k {\rm log} (u+\frac{\sigma}{2}) - k {\rm log}( u - \frac{\sigma}{2})}
\ee
Expanding the exponent, 
\ba
&&U(\sigma) = \frac{1}{\sigma}\oint \frac{du}{2\pi i}e^{- \sigma u^p}\nonumber\\
&&\times {\rm exp}[ - \frac{p(p-1)}{3! 4} \sigma^3 u^{p-2} - \frac{p(p-1)(p-2)(p-3)}{5! 4^2}
\sigma^5 u^{p-4} - \cdots]\nonumber\\
&&\times  [ 1 +  k (\frac{1 }{u} \sigma + \frac{1}{12 u^3}\sigma^{3} + \frac{1}{80 u^5}\sigma^{5} \cdots)
+ \frac{1}{2}k^2 (\frac{1}{u^2}\sigma^2 + \frac{1 }{6 u^4}\sigma^4 + \cdots)\nonumber\\
&&+ \frac{1}{3!}k^3 (\frac {1}{ u^3}\sigma^{3} + \cdots) + \frac{1}{4!}k^4 (\frac{1 }{ u^4} \sigma^4 +\cdots) + O(k^5)]
\ea
By choosing an integration path around the cut, with $x =\sigma u^p $, the above equation becomes
\ba\label{openp}
&&U(\sigma) =   \frac{1}{p \sigma^{1 + \frac{1}{p}}\pi }
\int_0^\infty dx x^{\frac{1}{p}-1} e^{-x} \nonumber\\
&&\times{\rm exp}[ - \frac{p(p-1)}{3!4}\sigma^{2+\frac{2}{p}}x^{1 - \frac{2}{p}} - 
\frac{p(p-1)(p-2)(p-3)}{5! 4^2 } \sigma^{4+\frac{4}{p}}x^{1-\frac{4}{p}}+ \cdots]\nonumber\\
&& [ 1 + k ( \sigma^{1 + \frac{1}{p}}x^{-\frac{1}{p}} + \frac{1}{12}\sigma^{3+\frac{3}{p}}x^{-\frac{3}{p}}+ \frac{1}{5\cdot 2^4}\sigma^{5+\frac{5}{p}}x^{-\frac{5}{p}}+\cdots) \nonumber\\
&&+ \frac{k^2}{2}(\sigma^{2+ \frac{2}{p}} x^{-\frac{2}{p}} + \frac{1}{6}\sigma^{4+\frac{4}{p}}x^{-\frac{4}{p}}+ \cdots) + \frac{1}{3!}k^3 (\sigma^{3+\frac{3}{p}}x^{-\frac{3}{p}}+\cdots)\nonumber\\
&& + \frac{1}{4!}k^4 (\sigma^{4+\frac{4}{p}}x^{-\frac{4}{p}}+\cdots) + ...]
\ea
The integration over $x$ gives
\ba
&&U(\sigma) = - (\frac{p-1}{24} + \frac{k^2}{2}) \frac{1}{\pi} \sigma^{1+\frac{1}{p}}\Gamma(1 - \frac{1}{p}) - (\frac{p}{24} k + \frac{1}{12}k^3) \frac{1}{\pi} \sigma^{2+ \frac{2}{p}}\Gamma( 1 - \frac{2}{p})\nonumber\\
&&- \frac{1}{144}[ - \frac{(p-1)(p-3)(1 + 2p)}{40} + (3 p +1) k^2 + 2 k^4] \frac{1}{\pi} \sigma^{3 + \frac{3}{p}}\Gamma( 1 - \frac{3}{p}) + ....\nonumber\\
\ea

This expansionc provides the following open intersection numbers, 
\ba\label{openinter}
&&<\tau_{1,0}>_{g=1} = \frac{p - 1 + 12 k^2}{24}
\nonumber\\
&&< \tau_{2,1}>_{g=\frac{3}{2}} = \frac{1}{24}(p k + 2 k^3)\hskip3mm (p\ne 2),\hskip 3mm
<\tau_{\frac{5}{2}}>_{g=\frac{3}{2}} = \frac{1}{12}(k + k^3) \hskip 3mm (p = 2)\nonumber\\
&& < \tau_{n,j}>_{g=2} =\frac{1}{p (12)^2} [\frac{(p-1)(p-3)(1+ 2 p)}{40} - (3p +1) k^2 - 2 k^4]\frac{\Gamma(1 - \frac{3}{p})}{\Gamma( 1 - \frac{1+ j}{p})}\nonumber\\
\ea
where the condition $(p+1)(2 g -1) = p n + j + 1$ determines $n$ and $j$. For $p$=2, $g=2$,
we have from the above expression,
\be
<\tau_{4,0}>_{g=2} = 
= \frac{1}{(24)^2 2!}[1  + 56 k^2 + 16 k^4]
\ee
which agrees with the result of  (\ref{tau4}).
The higher order open intersection numbers of $p$-th spin curves and  one marked point are
easily evaluated from the expansion of (\ref{openp}).
The intersection numbers are related to $U(\sigma)$ as in \cite{BHC3}
\be
U(\sigma) = \frac{1}{\pi}\sum <\tau_{n,j}>_g \Gamma(1 - \frac{1+j}{p}) p^{g-1}\sigma^{(2g-1)(1+\frac{1}{p})}
\ee
For $p=2$, (Kontsevich-Penner model), we have
\be
U(\sigma)= \frac{1}{2 \pi \sigma^{\frac{3}{2}}} e^{-\frac{1}{12}\sigma^3}\lim_{p\to 2}\int_0^\infty
dx e^{-x} x^{\frac{1}{p}-1}\left ( \frac{1 + \frac{1}{2}\sigma^{1 + \frac{1}{p}}x^{-\frac{1}{p}}}{1 - \frac{1}{2}\sigma^{1 + \frac{1}{p}}x^{-\frac{1}{p}}}\right )^k
\ee
Additional computations of open intersection numbers for $p$-spin curves are listed in an Appendix. In this Appendix, we derive also the string equation for $p$ spin curves in the presence of a logarithmic potential.

\vskip 3mm
\noindent{\bf{open $O(2N)$ model}}
\vskip 3mm

We now consider the non orientable intersection numbers provided by the  $O(2N)$ model. It is natural to investigate the relation between the non-orientable intersection numbers given by the $O(2N)$ model and the open intersection numbers which we have just discussed. The open intersection numbers for the $O(2N)$ case with a logarithmic potential is also interesting since the model deviates  from KdV and 
KP hierarchies.

For the O(2N) case with a logarithmic potential, $U(\sigma)$ for the $p$-th higher Airy singularity becomes (\ref{onepoint}),
\be
U(\sigma) = \frac{1}{N\sigma}\oint\frac{du}{2\pi i}e^{-\frac{c}{p+1}[(u+\frac{\sigma}{4})^{p+1}- (u - \frac{\sigma}{4})^{p+1}] + k {\rm log} (\frac{u + \frac{\sigma}{4}}{u - \frac{\sigma}{4}})}(1 - \frac{\sigma}{4u})
\ee
Since it resembles to the unitary case, with the replacement $\sigma $ by $2 s$,   the expansion (\ref{openp}) can be used.
\ba\label{openpO}
&&U(s) =   \frac{1}{p s^{1 + \frac{1}{p}}\pi } \int_0^\infty dx x^{\frac{1}{p}-1} e^{-x} ( 1 - \frac{1}{2}s^{1+\frac{1}{p}}x^{-\frac{1}{p}})
\nonumber\\
&&\times{\rm exp}[ - \frac{p(p-1)}{3!4}s^{2+\frac{2}{p}}x^{1 - \frac{2}{p}} - 
\frac{p(p-1)(p-2)(p-3)}{5! 4^2 } s^{4+\frac{4}{p}}x^{1-\frac{4}{p}}+ \cdots]\nonumber\\
&& [ 1 + k ( s^{1 + \frac{1}{p}}x^{-\frac{1}{p}} + \frac{1}{12}s^{3+\frac{3}{p}}x^{-\frac{3}{p}}+ \frac{1}{5\cdot 2^4}s^{5+\frac{5}{p}}x^{-\frac{5}{p}}+\cdots) \nonumber\\
&&+ \frac{k^2}{2}(s^{2+ \frac{2}{p}} x^{-\frac{2}{p}} + \frac{1}{6} s^{4+\frac{4}{p}}x^{-\frac{4}{p}}+ \cdots) + \frac{1}{3!}k^3 ( s^{3+\frac{3}{p}}x^{-\frac{3}{p}}+\cdots)\nonumber\\
&& + \frac{1}{4!}k^4 ( s^{4+\frac{4}{p}}x^{-\frac{4}{p}}+\cdots) + ...]
\ea
The term $( - \frac{1}{2}s^{1+\frac{1}{p}}x^{-\frac{1}{p}})$ gives
an additional contribution to the open intersection numbers characterized by a parameter $k$ as discussed in (\ref{openinter}). This contribution reads
\ba
U(s) &=& U_{0}(s) +\Delta U(s)\nonumber\\
\Delta U(s) &=& \frac{1}{\pi} [ \frac{k}{2}s^{1+\frac{1}{p}}\Gamma( 1 - \frac{1}{p}) + 
\frac{p-1}{48}s^{2 + \frac{2}{p}}\Gamma(1 - \frac{2}{p})\nonumber\\
&+& \frac{1}{72}( k + 2 k^3) s^{3+\frac{3}{p}} \Gamma( 1 - \frac{3}{p}) + \cdots ]
\ea
where $U_{0}(s)$ is the same as $U(\sigma)$ in (\ref{openp}). Thus the open intersection numbers for the $O(2N)$ case (  $O(2N)$ $p$-th Airy matrix model with a logarithmic potential),  together with $U_0(s)$, are given by 
\ba
&&<\tau_{1,0}>_{g=1} = \frac{p-1 + 12 k + 12 k^2}{24}\nonumber\\
&&<\tau_{2,1}>_{g=\frac{3}{2}} = \frac{( p-1) +  2 p k +  6 k^2 + 4 k^3  }{48},\hskip 3mm
<\tau_{\frac{5}{2}}>_{g=\frac{3}{2}}=\frac{1 + 4k + 6 k^2 + 4 k^3}{48}\nonumber\\
&& <\tau_{n,j}>_{g=2} = \frac{1}{p (12)^2}[ \frac{(p-1)(p-3)(1+ 2p)}{40} + 2 k 
-(1 + 3p) k^2 + 4 k^3 - 2 k^4] \nonumber\\
&&\times \frac{\Gamma(1-\frac{3}{p})}{\Gamma(1 - \frac{1 + j}{p})}
\ea
%%%%%%%%%%%%%%%%%%%%%%%%%%%%%%%%%
\vskip 3mm
\section{ Multiple marked points and Virasoro equations}
\setcounter{equation}{0}
\renewcommand{\theequation}{6.\arabic{equation}}
\vskip 5mm
\noindent{\bf{string equation}}
\vskip 2mm

The Virasoro equations have been investigated for the Kontsevich-Penner model \cite{BH07,Alexandrov2}. The first  Virasoro equation, or string equation,  reads   (\ref{string1}) and (\ref{string3}).  
\be\label{string5}
<\tau_0 \prod_{i=1}^s \tau_{n_i}>_g = \sum_{j=1}^s <\tau_{n_j - 1} \prod_{i\ne j}^s 
\tau_{n_i}>_g
\ee
Since the intersection numbers for $s$-marked points are known explicitly from the integral formula for$U(\sigma_1,...,\sigma_s)$, it is interesting to derive the above string equation and the other Virasoro equations for the
Kontsevich-Penner model from our  formulation of  the $s$-point correlation function $U(\sigma_1,...\sigma_s)$.
\vskip 2mm
 {\it{ $2$ marked points }}:
\vskip 3mm
The two marked points correlation function $U(\sigma_1,\sigma_2)$ is \cite{BHC3}
\ba
&&U(\sigma_1,\sigma_2) = e^{-\frac{1}{12} (\sigma_1^3+\sigma_2^3)}\oint \frac{du_1 du_2}{(2\pi i)^2} \exp \{ - \sigma_1 u_1^2 - \sigma_2 u_2^2 + k {\rm log}(\frac{u_1 + \frac{1}{2}\sigma_1}{u_1 - \frac{1}{2}\sigma_1})\nonumber\\
&& + k {\rm log}(\frac{u_2 + \frac{1}{2}\sigma_2}{u_2 - \frac{1}{2}\sigma_2})\}
 \frac{1}{(u_1 - u_2 + \frac{1}{2}(\sigma_1+ \sigma_2))(u_2 - u_1 +\frac{1}{2}(\sigma_1 + \sigma_2))}.
\ea
Writing the denominators as principal parts integrals as in  \cite{BHV}
\be\label{alpha}
\frac{1}{u_1- u_2 + \frac{1}{2}(\sigma_1+ \sigma_2)} = - i \int_0^\infty d\alpha e^{i \alpha
(u_1-u_2 + \frac{1}{2}(\sigma_1+ \sigma_2))}
\ee
we obtain
\ba
&&U(\sigma_1,\sigma_2) = e^{-\frac{1}{12} (\sigma_1^3+\sigma_2^3)}\oint \frac{du_1du_2}{(2\pi i)^2} \int_0^\infty d\alpha d\beta \exp \{- \sigma_1 u_1^2 - \sigma_2 u_2^2 \nonumber\\
&&+ k {\rm log}(\frac{u_1 + \frac{1}{2}\sigma_1}{u_1 - \frac{1}{2}\sigma_1})
 + k {\rm log}(\frac{u_2 + \frac{1}{2}\sigma_2}{u_2 - \frac{1}{2}\sigma_2}) + i(\alpha
- \beta)(u_1- u_2)\nonumber\\
&& + i (\alpha + \beta)\frac{1}{2}(\sigma_1+\sigma_2) \}
\ea
Replacing $u_i \to \frac{1}{\sqrt{\sigma_i}}$ and $\alpha\to \sqrt{\sigma_1\sigma_2}\alpha$ and $\beta\to \sqrt{\sigma_1\sigma_2}\beta$,
we obtain
\ba\label{U2}
&&U(\sigma_1,\sigma_2) = \sqrt{\sigma_1\sigma_2} e^{-\frac{1}{12}(\sigma_1^3+\sigma_2^3)}\oint \frac{du_1du_2}{(2\pi i)^2}\int_0^\infty d\alpha d\beta {\exp} \{ -u_1^2-u_2^2\nonumber\\
&+& i (\alpha-\beta)(\sqrt{\sigma_2}u_1 - \sqrt{\sigma_1}u_2) 
+  i \frac{1}{2}(\alpha + \beta)(\sqrt{\sigma_1\sigma_2}(\sigma_1+\sigma_2)\nonumber\\
&+& k {\rm log}(\frac{u_1+\frac{1}{2}\sigma_1^{3/2}}{u_1-\frac{1}{2}\sigma_1^{3/2}})
+ k {\rm log}(\frac{u_2+\frac{1}{2}\sigma_2^{3/2}}{u_2-\frac{1}{2}\sigma_2^{3/2}}) \}
\ea
Keeping the term of order $\sqrt{\sigma_2}$ in the pre factor and  putting the $\sigma_2$ to  0  in the exponent, we obtain
\be
U(\sigma_1,\sigma_2) = (\sqrt{\sigma_2}\oint \frac{du_2}{2\pi i} e^{-u_2^2}\frac{1}{u_2^2})
e^{-\frac{1}{12}\sigma_1^3}\oint \frac{du_1}{2\pi i}{\exp}[- \sigma_1 u_1^2 + k {\rm log}( \frac{u_1 + \frac{1}{2}\sigma_1}{u_1 - \frac{1}{2}\sigma_1})]
\ee
Using the argument of (\ref{even}) for the integration over $u_2$, we have for the one marked point
open intersection number $<\tau_{n-1}>_g$ for the Kontsevich-Penner model,
\be
<\tau_0 \tau_n>_g = <\tau_{n-1}>_g
\ee
which gives the results of the string equation (\ref{string5}) ; for instance,
\ba
&&<\tau_0 \tau_{\frac{1}{2}}>_{g=\frac{1}{2}} = <\tau_{-\frac{1}{2}}>_{g=\frac{1}{2}}
= k\nonumber\\
&&<\tau_0\tau_2>_{g=1}= <\tau_1>_{g=1}= \frac{1 + 12 k^2}{24}\nonumber\\
\ea
In Appendix, we show that the string equation holds also for the $p$ spin curves.
\vskip 3mm
{\it{ 3 marked points}}:
\vskip 3mm
The three point correlation function $U(\sigma_1,\sigma_2,\sigma_3)$ is given by
\ba
&&U(\sigma_1,\sigma_2,\sigma_3) = e^{-\frac{1}{12}(\sigma_1^3+\sigma_2^3+\sigma_3^3)}
\oint \prod_{i=1}^3\frac{du_i}{2\pi i} e^{-\sum \sigma_i u_i^2}\prod_{i=1}^3(\frac{u_i + \frac{1}{2}\sigma_i}{u_i - \frac{1}{2}\sigma_i})^k\nonumber\\
&&\frac{1}{[u_1-u_2+ \frac{1}{2}(\sigma_1+ \sigma_2)][u_2-u_3 + \frac{1}{2}(\sigma_2+\sigma_3)][u_3- u_1 + \frac{1}{2}(\sigma_3+ \sigma_1)]}\nonumber\\
\ea
The three denominators are replaced by  integrals ofver $\alpha,\beta,\gamma$ as in  (\ref{alpha}).
Changing variables  $u_i\to \frac{1}{\sqrt{\sigma_i}}$, $\alpha\to \sqrt{\sigma_1\sigma_3}\alpha$, $\beta\to \sqrt{\sigma_2\sigma_3}\beta$, $\gamma\to \sqrt{\sigma_3\sigma_1}\gamma$,  the above expression becomes
 \ba\label{U3}
 &&U(\sigma_1,\sigma_2,\sigma_3) = e^{-\frac{1}{12}(\sigma_1^3+\sigma_2^3+\sigma_3^3)}\sqrt{\sigma_1\sigma_2\sigma_3}
\oint \prod_{i=1}^3\frac{du_i}{2\pi i} e^{-\sum u_i^2}\prod_{i=1}^3(\frac{u_i + \frac{1}{2}\sigma_i^{3/2}}{u_i - \frac{1}{2}\sigma_i^{3/2}})^k\nonumber\\
&&\int_0^\infty d\alpha d\beta d\gamma e^{i\alpha(\sqrt{\sigma_2}u_1 - \sqrt{\sigma_1}u_2 + \frac{\sqrt{\sigma_1\sigma_2}}{2}(\sigma_1 +\sigma_2))}e^{i\beta(\sqrt{\sigma_3}u_2 - \sqrt{\sigma_2}u_3 + \frac{\sqrt{\sigma_2\sigma_3}}{2}(\sigma_2 +\sigma_3))}\nonumber\\
&&e^{i\gamma(\sqrt{\sigma_1}u_3 - \sqrt{\sigma_3}u_1 + \frac{\sqrt{\sigma_3\sigma_1}}{2}(\sigma_3 +\sigma_1))}
\ea
Keeping $\sqrt{\sigma_2}$, and putting  the other $\sigma_2$ to  zero, this expression becomes
\ba\label{U33}
&&\sqrt{\sigma_2}e^{-\frac{1}{12}(\sigma_1^3+\sigma_3^3)}\oint \frac{du_2}{2\pi i}
\frac{e^{-u_2^2}}{u_2^2}\oint \frac{du_1du_3}{(2\pi i)^2} \frac{e^{-u_1^2-u_3^2}}{\sqrt{\sigma_1}u_3 - \sqrt{\sigma_3}u_1 + \frac{1}{2}\sqrt{\sigma_1\sigma_3}(\sigma_1+\sigma_3)}\nonumber\\
&&(\frac{u_1+ \sigma_1^{3/2}}{u_1- \sigma_1^{3/2}})^k (\frac{u_3+ \sigma_3^{3/2}}{u_3- \sigma_3^{3/2}})^k\nonumber\\
&=&\sqrt{\sigma_2}e^{-\frac{1}{12}(\sigma_1^3+\sigma_3^3)}\oint \frac{du_2}{2\pi i}
\frac{e^{-u_2^2}}{u_2^2}\oint \frac{du_1du_3}{(2\pi i)^2} \frac{e^{- \sigma_1 u_1^2-\sigma_3 u_3^2}}{u_3 - u_1 + \frac{1}{2}(\sigma_1+\sigma_3)}\nonumber\\
&&\frac{u_1 - u_3 + \frac{1}{2}(\sigma_1+ \sigma_3)}{u_1 - u_3 + \frac{1}{2}(\sigma_1+ \sigma_3)}(\frac{u_1+ \sigma_1}{u_1- \sigma_1})^k (\frac{u_3+ \sigma_3}{u_3- \sigma_3})^k\nonumber\\
&=&\sqrt{\sigma_2}(\sigma_1+\sigma_3)U(\sigma_1,\sigma_3)
\ea
Thus we find a string equation for three marked points of the Kontsevich-Penner model,
\be
<\tau_0 \tau_{n_1}\tau_{n_2}>_g = <\tau_{n_1-1}\tau_{n_2}>_g + <\tau_{n_1}\tau_{n_2-1}>_g.
\ee
\vskip 3mm

Repeating the same procedure, we have  a string equation of $s$-marked points for the  Kontsevich-Penner model,
\be\label{stringequation}
<\tau_0\prod_{i=1}^s \tau_{n_i}>_g = \sum_{j=1}^s <\tau_{n_j-1}\prod_{i\ne j}^s\tau_{n_i}>_g
\ee

\vskip 3mm
\noindent{\bf{ W-constraints equation}}
\vskip 3mm
We consider next the terms of order $\sigma_2$ in the expression  {\it 2 marked points} $U(\sigma_1,\sigma_2)$.  The term $\sigma_2$ corresponds to $t_{\frac{1}{2}} \sim \tr \frac{1}{\Lambda}$. Such  fractional indices correspond to  $W$ constraints \cite{BH07}, which appear in the $p$-th higher Airy matrix model. Such a fractional index appears also in the non-orientable
Lie algebra $O(2N)$ as we have seen. Therefore, the terms $t_{n+ \frac{1}{2}}$ are characteristics of open intersection numbers. Keeping the  order $\sigma_2$ terms in (\ref{U2}), and putting the other
$\sigma_2$ to  zero,
\ba
&&U(\sigma_1,\sigma_2) = \sigma_2 \sqrt{\sigma_1}e^{-\frac{1}{12}\sigma_1^3}
\int_0^\infty d\alpha d \beta \oint \frac{du_1 du_2}{(2\pi i)^2} e^{- u_1^2 - u_2^2 + k {\rm log}(\frac{u_1+ \frac{1}{2}\sigma_1^{3/2}}{u_1- \frac{1}{2}\sigma_1^{3/2}})}\nonumber\\
&&e^{-i(\alpha + \beta)\sqrt{\sigma_1}u_2}
  (\alpha+ \beta) (i u_1)\nonumber\\
   &=& (2 \sigma_2\oint \frac{du_2}{2\pi i} \frac{e^{-u_2^2}}{u_2^3})\frac{1}{\sigma_1}e^{-\frac{1}{12}\sigma_1^3}\oint \frac{du_1}{2\pi i} u_1 e^{- u_1^2}\nonumber\\
&& (1 +  \frac{k\sigma_1^{\frac{3}{2}}}{u_1} + \frac{k^2 \sigma_1^3}{2 u_1^2}
  + \frac{(k + 2 k^3)\sigma_1^{\frac{9}{2}}}{u_1^3} + \cdots)
  \nonumber\\
  &=& 2 \sigma_2 \sigma_1^{\frac{1}{2}}R_3 R_0 k + \sigma_2\sigma_1^2 k^2 R_3 R_1
  + 2\sigma_2 \sigma_1^{\frac{7}{2}}R_3 (- k R_0 + (k + 2 k^3)R_2) + \cdots\nonumber\\
   \ea
   where we use the following contour integrals,
  \ba
&&R_{2n} = \oint \frac{du}{2\pi i} \frac{1}{u^{2n}}e^{- u^2} = \Gamma(\frac{1}{2}- n)\nonumber\\
&&R_{2n+1} = \oint \frac{du}{2\pi i} \frac{1}{u^{2n+1}}e^{- u^2} 
      = \lim_{p\to 2} \oint \frac{du}{2\pi i} \frac{1}{u^{2n+1}}e^{- u^p}
      = \lim_{p\to 2} \frac{2}{p}\Gamma(1 - \frac{2n}{p})\nonumber\\
&&R_2 = - 2 R_0 = -2 \Gamma(\frac{1}{2})
   \ea
From this expression, we have
\ba
< \tau_{\frac{1}{2}}\tau_{0}>_{g=\frac{1}{2}} = k,\hskip 3mm
<\tau_{\frac{1}{2}}\tau_{\frac{3}{2}}>_{g=1} =  k^2,\hskip 3mm
<\tau_{\frac{1}{2}}\tau_3>_{g=\frac{3}{2}} = \frac{1}{6}(3 k + 4 k^3)
\ea

\vskip 3mm
\noindent{\bf{dilaton equation}}
\vskip 3mm
Next, we consider the dilaton equation, which involve $\sigma_{\frac{3}{2}}$, i.e. $\tau_1$ in the intersection numbers.
The equation  $L_0 Z = 0 $ is a dilaton equation, which is derived by considering the terms of 
order $\tr \frac{1}{\Lambda^{\frac{3}{2}}}$.
For the dilaton equation, we have
\be\label{dilaton}
<\tau_{1}\prod_{i=1}^s \tau_{n_i}>_g = (2g - 2 + s) <\prod_{i=1}^s \tau_{n_i}>_g
\ee
From the three point function $U(\sigma_1,\sigma_2,\sigma_3)$, we obtain
\be
<\tau_0\tau_{\frac{1}{2}}\tau_1>_{g=\frac{1}{2}} = <\tau_0\tau_{\frac{1}{2}}>_{g=\frac{1}{2}}= k,\hskip 3mm <\tau_0^3 \tau_1>_{g=0} = <\tau_0^3>_{g=0} = 1, ...
\ee
which satisfies the dilaton equation of (\ref{dilaton}).
\vskip 3mm

For $\sigma_2^{\frac{3}{2}}$ in $U(\sigma_1,\sigma_2,...,\sigma_s)$,  scaling $u_i\to (\frac{x_i}{\sigma_i})^{\frac{1}{2}}$, it is obvious that the logarithmic term for $u_2$ can be neglected, since it gives higher orders.
Therefore, the dilaton equation does not show the effect of the logarithmic term and the equation is same as the dilaton equation without the logarithmic potential (there is no $k$ in the equation) as (\ref{dilaton}), and  thus we find  that (\ref{dilaton}) holds.

In Appendix, we discuss the string equation, the divisor equation and the dilaton equation for $p$ spin curves in the presence of the logarithmic potential.
\vskip 3mm
\noindent{\bf{Virasoro equations for Kontsevich-Penner model}}
\vskip 3mm
The Virasoro equations are expressed through operators $L_n$, which act on the partition function $Z = 
e^{F}$,
\be
L_n Z = L_n e^F = 0
\ee
with $n=-1, 0,1, ...$ ; for $n=-1$ it gives the string equation and for  $n=0$ the  dilaton equation.
The free energy is a generating function of the intersection numbers with a variables $t_n= \tr \frac{1}{\Lambda^{n+ \frac{1}{2}}}$.
For the Kontsevich-Penner model, since there is a logarithmic potential, we need to  consider also  half-integer values for $n$ in $t_{\frac{n}{2}}$. 
We recall  previous calculations \cite{BH07} and summarize here a  comparison with the calculation based$U(\sigma_1,...,\sigma_s)$. The first Virasoro equation for the order $\tr \frac{1}{\Lambda^{\frac{1}{2}}}$, which is the  string equation, is given by 
\be
\frac{\partial F}{\partial t_0} = \frac{1}{4}t_0^2 - \frac{k}{2}t_{\frac{1}{2}} + \sum_{n=0,\frac{1}{2},1, ...} (n + \frac{1}{2}) t_{n+1}\frac{\partial F}{\partial t_n}
\ee
The free energy is a generating function of the intersection numbers as
\be\label{coefficient}
F = \sum_{d_n} <\prod_n \tau_{n}^{d_n}> \prod_n \frac{\hat t_n^{d_n}}{d_n!}
\ee
where 
\be
\hat t_n = \tr \frac{1}{(2^{\frac{2}{3}}\Lambda)^{n+\frac{1}{2}}}
\ee
We have
the relation
\be
 t_n = (2^{\frac{2}{3}})^{n+ \frac{1}{2}} \hat t_n
 \ee
 
In \cite{BH07}, the Virasoro equations up to third order (i.e. up to the dilaton equation) are obtained as
\ba
&&(- \frac{\partial}{\partial t_0} + \frac{1}{4}J_{-2}^{(2)} - \frac{k}{2}t_{\frac{1}{2}}) g = 0\hskip 3mm (string \hskip1mm equation)\nonumber\\
&&(- 2 \frac{\partial}{\partial t_{\frac{1}{2}}} - k t_0 
- \frac{1}{16} t_{\frac{3}{2}} - \frac{k^2}{4}t_{\frac{3}{2}} - \frac{1}{12}J_{-4}^{(3)} +
 \frac{k}{4}J_{-4}^{(2)} - \frac{1}{2}J_{-1}^{(2)}) g= 0\nonumber\\
&&(-3 \frac{\partial}{\partial t_1} - \frac{1}{16} - \frac{3}{4}k^2 + k t_0 t_{\frac{1}{2}}- \frac{1}{4}J_0^{(2)} - \frac{1}{4}J_{-3}^{(2)})g = 0\hskip 3mm (dilaton \hskip 1mm equation)\nonumber\\
\ea
with
\ba
&& J_m^{(1)} = \frac{\partial}{\partial x_m} - m x_{-m},\hskip 3mm (m = ...,-2,-1,0,1,2,...)\nonumber\\
&&J_{m}^{(2)} = \sum_{i+j=m}: J_i^{(1)}J_j^{(1)}:\nonumber\\
&&=\sum_{i+j=m}\frac{\partial^2}{\partial x_i \partial x_j} + 2 \sum_{-i+j = m} i x_i \frac{\partial}{\partial x_j} + \sum_{-i-j=m}(ix_i)(jx_j)\nonumber\\
&&J_{m}^{(3)} = \sum_{i+j+k=m}:J_i^{(1)}J_j^{(1)}J_k^{(1)}:\nonumber\\
&&=\sum_{i+j+k=m}\frac{\partial^3}{\partial x_i\partial x_j\partial x_k}+ 3 \sum_{-i+j+k=-4}
\ea
where $:\cdots:$ means normal ordering, i.e. pulling the differential operator to the right. $x_n = \frac{1}{n}t_{\frac{n-1}{2}}$.
The solution of the Virasoro equations,  which includes  the string equation,   gives \cite{BH07}
\ba
F &=& \frac{1}{12}t_0^3 + \frac{1}{48}(1 + 12 k^2) t_1 + \frac{1}{2}k t_0 t_{\frac{1}{2}}  \\
&& + \frac{1}{24} t_0^3 t_1 + (\frac{1}{192} + \frac{1}{16} k^2) t_1^2 + \frac{k}{4}t_0 t_{\frac{1}{2}} t_1 + \frac{k}{24} {(t_{\frac{1}{2}})}^3 + \frac{1}{96}(1+ 12 k^2)t_0t_2\nonumber\\
&& + \frac{k}{4}t_0^2 t_{\frac{3}{2}} + \frac{1}{4} k^2 t_{\frac{1}{2}}t_{\frac{3}{2}}+ \frac{1}{6}(k + k^3) t_{\frac{5}{2}}+ \cdots 
\ea

From this expression, the intersection numbers , which are defined as (\ref{coefficient}) are obtained by changing $t_n$ to $\hat t_n$  for small genus,
\ba\label{intersectionnumber3}
&&<\tau_0^3>_{g=0} = 1,　\hskip 3mm <\tau_1>_{g=1} = \frac{1+ 12 k^2}{24},\hskip 3mm
<\tau_0\tau_{\frac{1}{2}}>_{g=\frac{1}{2}} = k, \nonumber\\
&&<\tau_0^3\tau_1>_{g=0} = 1, \hskip 3mm <\tau_1^2>_{g=1}= \frac{1+ 12 k^2}{24},\hskip 3mm
<\tau_0\tau_{\frac{1}{2}}\tau_1>_{g=\frac{1}{2}} = k,\nonumber\\
&&< \tau_{\frac{1}{2}}^3>_{g=\frac{3}{2}} = k,\hskip 3mm
< \tau_0\tau_2>_{g=1} = \frac{1+ 12 k^2}{24},\hskip 3mm
<\tau_0^2\tau_{\frac{3}{2}}>_{g=\frac{1}{2}} = k,\nonumber\\
&& <\tau_{\frac{1}{2}}\tau_{\frac{3}{2}}>_{g=1}
= k^2,\hskip 3mm <\tau_{\frac{5}{2}}>_{g= \frac{3}{2}} = \frac{1}{12}(k + k^3)
\ea
These intersection numbers satisfy the string equation, the W-constraints, and  the dilaton equation. They provide identical  values  as those calculated from $U(\sigma_1,...,\sigma_s)$ for  the Kontsevich-Penner model, as shown in Appendix.
%%%%%%%%%%%%%%%%%%%%%%%%%%%%%%%%%%%%%%%
\vskip 3mm
\section{Gromov-Witten invariants of $CP^1$ model}
\setcounter{equation}{0}
\renewcommand{\theequation}{7.\arabic{equation}}
\vskip 5mm
The Gromov-Witten invariants of  $CP^1$ model has been studied \cite{Eguchi,Okounkov,Aganagic}. Recently, the
Gromov-Witten invariants are evaluated in more higher orders \cite{Norbury}.  We apply the present method to Gromov-Witten invariants of
$CP^1$ model, since it has a similar matrix model representation as Kontsevich type of the external source \cite{Aganagic}.

The $CP^1$ matrix model is described as
\be\label{CP1}
Z = \int D(e^M) e^{\tr( e^M + q e^{- M}) + \tr M \Lambda }
\ee
We use the Gaussian random matrix model with an external source, and by the tuning the external source $A$, we obtain (\ref{CP1}) as a generalized Kontsevich model. Therefore,
as before, we consider the Fourier transform of the density correlation functions. Particularly,
we consider $U(\sigma)$, which is
\ba
U(\sigma) &=& \frac{1}{\sigma}\oint \frac{du}{2\pi i} e^{e^{u+\frac{1}{2}\sigma} 
- e^{u - \frac{1}{2}\sigma}+ q e^{-(u+\frac{1}{2}\sigma)}- q e^{-(u-\frac{1}{2}\sigma)}+ u}\nonumber\\
&=& \frac{1}{\sigma}\oint\frac{du}{2\pi i} e^{(2\rm sinh\frac{\sigma}{2})(e^u - q e^{-u}) + u}
\ea
By $x = e^u$, we have
\be
U(\sigma) = \frac{1}{\sigma}\oint \frac{dx}{2\pi i}e^{(2N{\rm sinh}\frac{\sigma}{2N})(x - q x^{-1})}
\ee
where we inserted $N$ to make clear the genus expansion.
The residue calculation becomes
\be
U(\sigma)  \frac{1}{\sigma}\sum_{d=1}\frac{1}{d! (d-1)!}(2N{\rm sinh}\frac{\sigma}{2N})^{2d-1}(-q)^d
\ee
In the genus zero, $N\to \infty$, dropping the irrelevant factor of $q$, we have
\ba
U(\sigma) &=& \sum_{d=1}^\infty \frac{1}{d!(d-1)!} \sigma^{2(d-1)}\nonumber\\
&=& \sum_{d=0}^\infty \frac{1}{(d+1)! d!} \sigma^{2d} = \sum_{d=0}^\infty <\tau_{2d}>_{g=0} (d+1) \sigma^{2d}
\ea
with 
\be
<\tau_{2d}>_{g=0} = \frac{1}{(d+1)!^2}.
\ee

For the higher genus $g$, we expand $(2N{\rm sinh}\frac{\sigma}{2N})^{2d-1}$, and pick up the
genus $g$ terms from order $\frac{1}{N^{2g}}$ terms.
\be
\frac{1}{\sigma}(2N{\rm sinh}\frac{\sigma}{2N})^{2d-1} = \sigma^{2d-2} + \frac{2d -1}{24 N^2} \sigma^{2d} + \frac{(2 d - 1)(10 d - 7)}{5760 N^4}\sigma^{2d + 2}
+ \cdots
\ee
With  the shift of power of $\sigma$, $2d+ 2n \to 2d$, we obtain Gromov-Witten invariants of genus $g$ as
\be
U(\sigma) = \sum_d \sum_g <\tau_{2d}>_g ( d + 1 - g) \sigma^{2d}
\ee
with 
\ba
&&<\tau_{2d}(\omega)>_{g=0} = \frac{1}{((d+1)!)^2}\nonumber\\
&&<\tau_{2d}(\omega)>_{g=1} = \frac{2d -1}{24 (d!)^2}\nonumber\\
&&<\tau_{2d}(\omega)>_{g=2} = \frac{d^2 ( 2 d - 3)(10 d - 17)}{2^7\cdot 3^2\cdot 5 (d!)^2}\nonumber\\
&&<\tau_{2d}(\omega)>_{g=3} = \frac{d^2 (d-1)^2  (2 d-5)(140 d^2 - 784 d + 1101)}{2^{10}\cdot 3^4\cdot 5\cdot 7 (d!)^2}
\ea
These numbers agree with the result of the recent evaluation by Norbury and Scott by a different method up to genus three\cite{Norbury}.
It is straight forward to evaluate Gromov-Witten  one point invariants in any order of genus from $U(\sigma)$. 
We have for $g=4$ as
\be
<\tau_{2d}(\omega)>_{g=4} =\frac{d^2 (d-1)^2 (d-2)^2}{(d!)^2} \frac{(2 d-7)(10 d-39)(140 d^2 -1092 d + 2143)}{2^{15}\cdot 3^5 \cdot 5^2\cdot 7}
\ee

   %%%%%%%%%%%%%%%%%%%%%%%%%%%%%%%%%
\vskip 3mm
\section{Discussions}
\setcounter{equation}{0}
\renewcommand{\theequation}{9.\arabic{equation}}
\vskip 5mm
 In this article, we have considered the generalization of the Airy matrix model to a  $p$-th singularity. This   provides the intersection numbers of the moduli space of $p$-spin curves for  orientable and non-orientable Riemann surfaces, with Lie algebras of  $U(N)$, $O(2N)$, $O(2N+1)$ and $Sp(N)$.  The Euler characteristics are easily evaluated  by taking  the $p\to -1$ limit.  Our results are consistent with the two categories, orientable and non-orientable surfaces, since we have obtained two type of topological invariants (two different Euler characteristics) for Lie algebras. The  expressions agree with the virtual Euler characteristics obtained earlier \cite{Goulden} for non-orientable surfaces. 
 We have obtained explicit expressions to all order in the genus for one marked point in the $p=3$ and $p=4$ cases  given in terms of  Bessel functions.  
 
  For the open intersection numbers, which are defined by the insertion of a disk on a closed Riemann surface as a boundary, we have used the  Kontsevich-Penner model. We have derived the  Virasoro equations, string equation and dilaton equations,  for this Kontsevich-Penner model from explicit integral representations. The open intersection numbers are extended to $p$--spin curves,  from a  higher Airy matrix model with a logarithmic potentials.
  
   In our 
  previous article \cite{BHC3}, the Airy matrix model with a logarithmic potential  was derived from  the average of two characteristic polynomials  in a  two matrix model with an external source The eigenvalues of the first matrix  $M_1$ is on an edge of the distribution, and  for the other matrix $M_2$ in the bulk,. After integration  over the  matrix $M_2$, a model with a logarithmic potential is obtained for $M_1$. The coefficient of the logarithmic potential $k$ corresponds to the power of $({\rm det} M_1)^k$. This logarithmic potential provides  the boundary for the open intersection theory. 
  
    The integral representation of  the $s$ point correlation function for a Gaussian matrix model with an external source provides a powerful tool for the evaluation of open/close intersection numbers and Gromov-Witten invariants. It would be interesting to extend the present analysis to  more complicated cases ,  such as the  Gromov-Witten theory of $CP^{n-1}$.
  
\vskip 10mm
{\large{\bf Acknowledgements}}
\vskip 2mm
S.H. thanks  the support  of 
JSPS KAKENHI Grant Number 25400414 .

\newpage
{\large{\bf Appendix:  Virasoro equations of open intersection numbers for $p$ spin curves }}
\setcounter{equation}{0}
\renewcommand{\theequation}{A.\arabic{equation}}
\vskip 5mm

\vskip 3mm
The open/close intersection numbers of $p$-spin curves are evaluated from the expressions for the $s$-point
correlation functions $U(\sigma_1,...,\sigma_s)$. From (\ref{openp}), the one marked point intersection numbers $<\tau_{n,j}>_g$ are obtained as
\vskip 2mm
\ba
&&< \tau_{1,0}>_{1} = \frac{p - 1 + 12 k^2}{24}\nonumber\\
&&<\tau_{n,j}>_{\frac{3}{2}} = \frac{1}{24}(p k + 2 k^3) \frac{\Gamma(1 - \frac{2}{p})}{\Gamma(1 - \frac{1+j}{p})}\nonumber\\
&&<\tau_{n,j}>_2= \frac{1}{p(12)^2}\{\frac{(p-1)(p-3)(1 + 2p)}{40} - (1 + 3p) k^2 - 2 k^4\}\frac{\Gamma(1 - \frac{3}{p})}{\Gamma(1 - \frac{1+j}{p})}\nonumber\\
&&<\tau_{n,j}>_{\frac{5}{2}}= \{\frac{k}{5760} (18-25p + 30 p^2 - 5 p^3) + \frac{k^3}{144} (p-1)\}\frac{\Gamma(1 - \frac{4}{p})}{\Gamma(1 - \frac{1+j}{p})}\nonumber\\
&&<\tau_{n,j}>_{3}= \frac{1}{p^2}\{ \frac{1}{2903040}(p-1)(p-5)(1+2p)(8p^2-13 p - 13)\nonumber\\
&&\hskip 5mm + \frac{1}{57600}(-10 p^3+85 p^2+ 90 p+19) k^2 + \frac{1}{2880} (5p+3) k^4 + \frac{1}{3600} k^6 \}\nonumber\\
&&\hskip 5mm\times \frac{\Gamma(1 - \frac{5}{p})}{\Gamma(1 - \frac{1+j}{p})}
\nonumber\\
\ea
where $n$ and $j$ are constrained by the condition,
\be\label{s1}
(p+1)(2 g - 1) = p n + j + 1
\ee
i.e. he $s=1$ case of  the  general condition,
\be\label{sp}
(p+1) (2 g - 2 + s) = p \sum_{k=1}^s n_i + \sum_{k=1}^s j_{k} + s
\ee
For $<\tau_{\frac{5}{2}}>_{g=\frac{3}{2}}$ of $p=2$ is obtained from $\lim_{p\to 2} <\tau_{2,1}>_{g=\frac{3}{2}}$ since the right hand side of (\ref{s1}) is the same, and it becomes
$\frac{1}{12}(k + k^3)$.

For the 2 marked points, the intersection numbers of the  $p$ spin curves are derived from $U(\sigma_1,\sigma_2)$.
\ba
U(\sigma_1,\sigma_2) &=& \oint \frac{du_1du_2}{(2\pi i)^2} e^{- \frac{1}{p}\sum\limits_{i=1}^2 [(u_i + \frac{1}{2}\sigma_i)^{p+1}-(u_i - \frac{1}{2}\sigma_i)^{p+1}]}\prod_{i=1}^2 (\frac{u_i + \frac{1}{2}\sigma_i}{u_i - \frac{1}{2}\sigma_i})^k\nonumber\\
&&\frac{1}{(u_1- u_2 + \frac{1}{2}(\sigma_1+\sigma_2))(u_2- u_1 + \frac{1}{2}(\sigma_1+\sigma_2))}
\ea
\vskip 3mm
\noindent{\bf{string equation}}
\vskip 3mm
Using $x_i = \sigma_i u_i^2$, and $\alpha \to (\sigma_1\sigma_2)^{\frac{1}{p}}\alpha$, $\beta \to (\sigma_1\sigma_2)^{\frac{1}{p}}\beta$, taking the same process as (\ref{U2}), we obtain 
\ba\label{x12}
&&U(\sigma_1,\sigma_2) = \frac{(\sigma_1 \sigma_2)^{\frac{1}{p}}}{p^2 \pi^2}
\int_0^\infty \cdots \int_0^\infty dx_1 dx_2 d\alpha d\beta (x_1 x_2)^{\frac{1}{p}-1}e^{-x_1-x_2}\nonumber\\
&&e^{-\frac{p(p-1)}{24}\sum_i (\sigma_i^{2+\frac{2}{p}}x_i^{1-\frac{2}{p}})+ \cdots}\nonumber\\
&& e^{i\alpha((\sigma_2 x_1)^{\frac{1}{p}} - (\sigma_1 x_2)^{\frac{1}{p}} + \frac{1}{2}(\sigma_1\sigma_2)^{\frac{1}{p}}(\sigma_1+ \sigma_2))}
e^{i\beta((\sigma_1 x_2)^{\frac{1}{p}}- (\sigma_2 x_1)^{\frac{1}{p}} + \frac{1}{2}(\sigma_1\sigma_2)^{\frac{1}{p}}(\sigma_1+ \sigma_2))}\nonumber\\
&&\prod_{i=1}^2(\frac{x_i^{\frac{1}{p}}+ \frac{1}{2}\sigma_i^{1+ \frac{1}{p}}}{x_i^{\frac{1}{p}}- \frac{1}{2}\sigma_i^{1+ \frac{1}{p}}})^k
\ea
Taking the term of
order $\sigma_2^{\frac{1}{p}}$ and neglecting higher order terms in  $\sigma_2$, we have
\ba
U(\sigma_1,\sigma_2) &=& - \frac{\sigma_2^{\frac{1}{p}}}{\pi}\Gamma(1 - \frac{1}{p})\cdot
\oint \frac{du_1}{2\pi i} e^{- \frac{1}{p} [(u_1 + \frac{1}{2}\sigma_1)^{p+1}-(u_1 - \frac{1}{2}\sigma_1)^{p+1}]} (\frac{u_1 + \frac{1}{2}\sigma_1}{u_1 - \frac{1}{2}\sigma_1})^k
\nonumber\\
&=& - \frac{\sigma_2^{\frac{1}{p}}}{\pi}\Gamma(1 - \frac{1}{p})\cdot
\sigma_1 U(\sigma_1)
\ea
This equation is a string equation,
\be
<\tau_{0,0}\tau_{n,j}>_g = <\tau_{n-1,j}>_g
\ee
A string equation for three marked point for $p$ spin curves is an extension of (\ref{U33}). It is easily obtained from
 \be
 U(\sigma_1,\sigma_2,\sigma_3) = \sigma_2^{\frac{1}{p}}\Gamma(1 - \frac{1}{p})
 (\sigma_1+\sigma_3) U(\sigma_1,\sigma_3)
 \ee
 which is the  string equation,
 \be
 <\tau_{0,0}\tau_{n_1,j_1}\tau_{n_2,j_2}>_g = <\tau_{n_1-1,j_1} \tau_{n_2,j_2}>_g + 
 <\tau_{n_1,j_1} \tau_{n_2-1,j_2}>_g
 \ee
 \vskip 3mm
 \noindent{\bf{W constraint equation}}
 \vskip 3mm
 We consider next the W constraint equation.  Since there are spin $j=0,1,...,p-1$ indices
 for the intersection numbers of $p$ spin curves, we have an equation which involves
 $\tau_{0,j}$. Taking next $\sigma_2^{\frac{2}{p}}$, and neglecting higher terms in  $\sigma_2$ in (\ref{x12}), we obtain the intersection number with  $\tau_{0,1}$.
 
 The two point correlation function $U(\sigma_1,\sigma_2)$ is expressed as \cite{BH5},
 by using the following representation.
 \ba
&& \frac{1}{u_1 - u_2 + \frac{1}{2}(\sigma_1+\sigma_2)}\frac{1}{u_1-u_2 - \frac{1}{2}(\sigma_1+\sigma_2)}\nonumber\\
&& = \frac{1}{\sigma_1+\sigma_2}\int_0^\infty d\alpha e^{- \alpha (u_1 - u_2)}{\rm sinh}
 (\frac{\alpha}{2}(\sigma_1+\sigma_2))
 \ea
 By $\alpha\to (\sigma_1\sigma_2)^{\frac{1}{p}}\alpha$, $u_i = (x_i/\sigma_i)^{\frac{1}{p}}$, ${\rm sinh}(\frac{\alpha}{2}(\sigma_1+ \sigma_2)) \sim \frac{\alpha}{2}(\sigma_1\sigma_2)^{\frac{1}{2}}(\sigma_1+ \sigma_2)$, we obtain $\sigma_1^{\frac{1}{p}}$ term as
 \ba
 &&U(\sigma_1,\sigma_2) = \frac{1}{p^2}\frac{(\sigma_1\sigma_2)^{\frac{1}{p}}}{2}\int dx_1 dx_2 \frac{1}{(\sigma_2 x_1)^{\frac{2}{p}}}(x_1x_2)^{\frac{1}{p}-1}e^{- x_1 - x_2+ \cdots}\nonumber\\
 &&= \sigma_1^{\frac{1}{p}}\sigma_2 U(\sigma_2)
 \ea
 which is a string equation. For $\sigma_1^{\frac{2}{p}}$ term, we expand $e^{- (\sigma_1 x_2)^{\frac{1}{p}}\alpha}= 1 + (\sigma_1 x_2)^{\frac{1}{p}} \alpha$.
 \be
 U(\sigma_1,\sigma_2) = \frac{1}{p} \sigma_1^{\frac{1}{p}}\Gamma(1 - \frac{2}{p})
 \sigma_2^{- \frac{2}{p}}\int_0^\infty  dx_2 x_2^{\frac{2}{p}-1} e^{- x_2 - \frac{p(p-1)}{24}\sigma_2^{2 + \frac{2}{p}} x_2^{1 - \frac{2}{p}} + \cdots} (\frac{x_2^{\frac{1}{p}}+ \frac{1}{2} \sigma_2^{1 + \frac{1}{p}}}{x_2^{\frac{1}{p}} - \frac{1}{2} \sigma_2^{1 + \frac{1}{p}}})^k
 \ee
 From above integral, we obtain $\sigma_1^{\frac{2}{p}}\sigma_2^{4+ \frac{2}{p}}$ term as
 \be
 U(\sigma_1,\sigma_2) \sim t_{0,1}t_{4,1} [\Gamma(1-\frac{2}{p})]^2\{
 (1-\frac{2}{p})\frac{1}{2}\frac{p(p-1)^2}{(24)^2} - \frac{(p-1)(p-2)(p-3)}{5! 4^2}\}
 \ee
 Separating a factor $(1 - \frac{2}{p})$, we obtain in the case $k=0$,
 \be
 <\tau_{0,1}\tau_{4,1}>_g = <\tau_{3,2}>_g + \frac{1}{2p}<\tau_{1,0}>_g^2
 \ee
 for general $p$. Similarly we obtain for $g=3$, when $k=0$,
 \be
 <\tau_{0,1}\tau_{6,3}>_g = <\tau_{5,4}>_{g=3} + \frac{1}{p}<\tau_{1,0}>_{g=1}<\tau_{3,2}>_{g=2}
 \ee
 
 \vskip 2mm
 \noindent{\bf{dilaton equation}}
 \vskip 2mm
 The dilaton equation for $p$-spin curves is
 \be
 <\tau_{1,0}\prod_{k=1}^s \tau_{n_k,j_k}>_g = ( 2 g - 2 + s)<\prod_{k=1}^s \tau_{n_k.j_k}>_g
 \ee
  We consider $s=1$, two point correlation function $U(\sigma_1,\sigma_2)$.
 By the shift $\alpha\to (\sigma_1 \sigma_2)^{\frac{1}{p}}$, $x_i = \sigma_i u_i^p$,
 we have
 \ba
 U(\sigma_1,\sigma_2) &=& \frac{1}{\sigma_1+\sigma_2}\int_0^\infty d\alpha {\rm sinh}
 (\frac{\alpha}{2}(\sigma_1\sigma_2)^{\frac{1}{2}}(\sigma_1+\sigma_2))e^{-[(\sigma_2 x_1)^{\frac{1}{p}}-(\sigma_1 x_2)^{\frac{1}{p}}]+ O(\sigma^{2+ \frac{2}{p}})}\nonumber\\
 &&
 \prod (\frac{x_i+\frac{1}{2}\sigma_i^{\frac{3}{2}}}{x_i-\frac{1}{2}\sigma_i^{\frac{3}{2}}})^k
 \ea
 For simplicity, we evaluate the  $p=2$ case.
 The term of order $\sigma_1^{\frac{3}{2}}$ comes from
 \be
 \frac{1}{\sigma_1+\sigma_2}{\rm sinh}(\frac{\alpha}{2}(\sigma_1\sigma_2)(\sigma_1+\sigma_2))e^{\alpha(\sigma_1 x_2)^{\frac{1}{2}}}\sim
 \frac{1}{4}\alpha^3 \sigma_1^{\frac{3}{2}}\sigma_2^{\frac{1}{2}}x_2 + \frac{\alpha^3}{48}\sigma_1^{\frac{3}{2}}\sigma_2^{\frac{7}{2}}
 \ee
 By the integrations of $\alpha$ and $x_1$, we obtain the order of $\sigma_1^{\frac{3}{2}}$
 \be
 U(\sigma_1,\sigma_2) \sim
 2 \sigma_1^{\frac{3}{2}}\Gamma(1 - \frac{1}{2})[ \sigma_2^{-\frac{3}{2}}\int dx_2 x_2^{\frac{1}{2}}e^{- x_2+ \cdots} + \frac{1}{12}\sigma_2^{\frac{3}{2}}\int dx_2 x_2^{-\frac{1}{2}}e^{-x_2+\cdots} ]
 \ee
 Noting the integral by parts for the first term, we have
 \be\label{dilaton2}
 U(\sigma_1,\sigma_2) = 2 \sigma_1^{\frac{3}{2}}\Gamma(1 - \frac{1}{2}) ( 1 + \frac{1}{6}\sigma_2^3)U(\sigma_2)
 \ee
 Since we have
 \be
 U(\sigma_2) = \sum_g \frac{1}{(12)^g g!} (-1)^g \sigma_2^{3g - \frac{3}{2}}
 \ee
 we resum the two terms of (\ref{dilaton2}) as
 \be
 U(\sigma_1,\sigma_2) = 
 2\sigma_1^\frac{3}{2}\Gamma(\frac{1}{2}) (\sum_{g=1}^\infty \frac{(-1)^g}{(12)^g g!}\sigma_2^{3 (g-\frac{1}{2})}+ \sum_{g=1}^\infty \frac{(-1)^{g+1}}{(12)^g g!}\frac{1}{6}\sigma_2^{3(g+ \frac{1}{2})})
 \ee
  which provides the dilaton equation for $p=2$,
 \be
 <\tau_{1,0}\tau_{n,j}>_g = ( 2g -1) <\tau_{n,j}>_g
 \ee
 We can check , for instance, $g=1$ case for $p=2$ as
 \be
 <\tau_{1,0}^2>_{g=1} = <\tau_{1,0}>_{g=1} = \frac{1}{24}( 1 + 12 k^2)
 \ee
 	The above equation may be easily extended to  $p > 2$ by the same process.

 \newpage
%%%%%%%%%%%%%%%%%%%%%%%

\end{document}